\documentstyle[aas2pp4]{article}

\received{August 21, 1997}
\accepted{October 10, 1997}
\journalid{}{}
\articleid{}{}

\slugcomment{to appear in the Astronomical Journal (January 1998 issue)}

\lefthead{Kissler-Patig et al.}
\righthead{Keck Spectroscopy of Globular Clusters around NGC 1399}

\begin{document}
\input{psfig.sty}

\title{Keck Spectroscopy of Globular Clusters around NGC 1399$^1$}

\author{Markus Kissler-Patig, Jean P. Brodie, Linda L. Schroder}
\affil{UCO/Lick observatory, University of California,
    Santa Cruz, CA 95064}
\affil{Electronic mails: mkissler@ucolick.org, brodie@ucolick.org,
linda@ucolick.org}

\author{Duncan A. Forbes}
\affil{School of Physics and Astronomy, University of Birmingham,
Edgbaston, Birmingham B15 2TT, United Kingdom}
\affil{Electronic mail: forbes@star.sr.bham.ac.uk}

\and

\author{Carl J. Grillmair}
\affil{Jet Propulsion Laboratory, 4800 Oak Grove Drive, Pasadena, CA 91109}
\affil{Electronic mail: carl@bb1.jpl.nasa.gov}

\author{John P. Huchra}
\affil{Havard--Smithsonian Center for Astrophysics, 60 Garden Street,
Cambridge, MA 02138}
\affil{Electronic mail: huchra@fang.harvard.edu}

\altaffiltext{1}{Based on observations obtained at the W.~M.~Keck Observatory,
which is operated jointly by the California Institute of Technology and
the University of California.}

\begin{abstract}

We report moderate resolution, high signal--to--noise spectroscopy of globular 
clusters around NGC 1399, the central cD galaxy in the Fornax cluster. 
We address issues as diverse as element abundances of globular clusters
versus stellar populations in ellipticals, blue horizontal branches in
metal--rich globular clusters, broad--band colors as metallicity tracers,
possible overestimation of the age--metallicity degeneracy in globular
clusters, and dark matter in the halo of NGC 1399.

We obtained spectra for 21
globular cluster candidates with multi-slit spectroscopy using the LRIS 
on the KECK 1 telescope. Our sample turned out to include 18 globular
clusters, 1 star, and 2 low redshift late-type galaxies ($z\simeq0.3$).

The mean velocity of our globular cluster sample is $1293 \pm 71$ 
km$\cdot$s$^{-1}$ and its velocity dispersion $302 \pm 51$ km$\cdot$s$^{-1}$.
Both are slightly lower than, but in agreement with, previously derived values.
We derive a mass of 1 to $5\times 10^{12} M_\odot$ within 28 kpc for the galaxy,
and a $M/L_{\rm B}$ ratio of $36\pm20$ or $76\pm 40$ $M_\odot/L_{\odot \rm B}$,
depending on the mass estimator. Both estimates indicate that dark matter
dominates the potential at 6 r$_{\rm eff}$.

The derived element abundances for the globular clusters span the entire
range observed in the Milky Way and M31, with a mean metallicity of our
sample around $[$Fe/H$] \simeq -0.8$ dex. This implies that the two major
sub--populations reported from photometry could have formed by the same
processes as the ones that formed halo and disk/bulge globular clusters in
the Local Group spirals. Two globular clusters (that we associate 
with a group of very red globular clusters, representing about 5\% of the 
total system) clearly stand out and exhibit metal abundances 
as high as observed for stellar
populations in giant ellipticals. In addition, they display surprisingly high 
H$\beta$ and H$\gamma$
indices that are not explained by any age/metallicity combination of
existing models. 
The high Mg and H$\beta$ values in these clusters could, however,
be explained by the presence of blue horizontal branches.

Finally, we find that V$-$I and metallicity are well correlated in the 
globular cluster system, but that the slope of the relation is twice as flat 
at high metallicities as an extrapolation from the relation for Milky Way 
globular clusters. This implies that the mean metallicities of globular cluster
systems in ellipticals are
lower, and cover a smaller range, than previously derived from broad--band
V$-$I colors.

\end{abstract}


%%%%%%%%%%%%%%%%%%%%%%%%%%%% INTRO %%%%%%%%%%%%%%%%%%%%%%%%%%%%%%%%%%%
%%%%%%%%%%%%%%%%%%%%%%%%%%%%%%%%%%%%%%%%%%%%%%%%%%%%%%%%%%%%%%%%%%%%%

\section{Introduction}

Globular cluster systems were extensively studied using photometry in the
last two decades producing many interesting findings 
(e.g.~Ashman \& Zepf 1997 for a review). However, many new questions were raised
that could not be
answered using broad--band colors alone because population synthesis models
predict (sometimes strong) age--metallicity degeneracies (e.g.~Worthey
1994). Several attempts were made to obtain spectra of globular clusters
in systems outside the Local Group (Mould et al.~1987, Mould et al.~1990,
Huchra \& Brodie 1987, Brodie \& Huchra 1991, Grillmair et al.~1994,
Perelmuter, Brodie \& Huchra 1995, Hui et al.~1995, Bridges et al.~1997,
Minniti et al.~1998). However, all of these studies had to focus on the very 
brightest globular clusters in the host galaxies and only low signal--to--noise
spectra were obtained. Kinematic information could be obtained for the 
globular 
clusters but, for the vast majority, line indices were either affected by high 
photon noise, or not measurable at all.

Using the new generation of 10m--class telescopes, the expectation was that
reliable line indices would be measurable for many globular clusters in galaxies 
at typical
distances of the Virgo or Fornax galaxy clusters.
Among the open questions were: Do globular clusters in other galaxies show
the same abundances and abundance ratios as in the Local Group, i.e.~can
they have formed by similar processes? Are globular clusters in ellipticals
more metal rich than in spirals as thought from their broad--band colors
(e.g~first pointed out by Cohen 1988),
and what is the upper metallicity limit? Can the abundances help to
identify a population of globular clusters in ellipticals that formed
during a merger event? How do the abundances of globular clusters relate to
those of the stars of the host galaxies? 
Preliminary answers to some of these questions were discussed 
by Brodie \& Huchra (1991), but conclusive statements require better data
than was available to them.
 
Here we report the first spectroscopically derived abundances for globular 
clusters in NGC 1399, the central giant elliptical galaxy in Fornax.
NGC 1399, like other cD galaxies, has an over--abundant globular cluster
system with $\simeq 5800$ globular clusters (Kissler-Patig et al.~1997,
Forbes et al.~1997). Photometric studies suggest the presence of multiple 
globular cluster sub--populations (e.g.~Ostrov et al.~1993, Kissler-Patig
et al.~1997, Forbes et al.~1997), the origins of which are uncertain.
In the next section we describe the observations. In section 3 we derive
the kinematics of our globular cluster sample and use this information
 to give an estimate of the galaxy mass and mass--to--light ratio at several 
effective 
radii. The abundances and metallicities of our globular clusters  are
derived in Sect.~4. In section 5 we discuss our broad--band colors together
with the line indices. We draw conclusions from our sample in
Sect.~6.

%%%%%%%%%%%%%%%%%%%%%%%%%%%%  OBS %%%%%%%%%%%%%%%%%%%%%%%%%%%%%%%%%%%
%%%%%%%%%%%%%%%%%%%%%%%%%%%%%%%%%%%%%%%%%%%%%%%%%%%%%%%%%%%%%%%%%%%%%

\section{Observations and Reduction}

We selected 21 globular cluster candidates from the original list used
by Grillmair et al.~(1994) and kindly provided by W. Couch. These cluster 
candidates were identified on deep photographic plates in the Anglo-Australian 
Telescope archive and were selected on the basis of their B$_j-$R colors.

The observations were carried out with the Low Resolution Imaging
Spectrograph (LRIS, Oke et al.~1995) in multi-slit mode at the
Keck 1 telescope on the nights of 1995, December 22nd and 23rd. 
Exposures totaling 160 min were taken over the two nights for one mask
with 21 slitlets + 3 centering objects. Comparison lamps and flat
fields were taken after the 30 min + 20 min exposures on the night of the 
22nd, and before and after the 20 min + $3\times30$ min exposures in the night 
of the 23rd.   

The CCD used was a Tektronix 2048$\times$2048 with 24 $\mu$m pixels. The
observations were made with a 600g/mm grating (blazed at 5000 \AA, with a
dispersion 1.24\AA /pixel) and 1 arcsec slitlets, resulting in an
effective resolution of 5.6\AA\, and a usable wavelength range of 4000\AA\ to
6100\AA\ common to all spectra. 

The reduction was done in a standard fashion under IRAF, with the help of 
Kelson's et al.~(1997) EXPECTOR program.
Every exposure was bias--corrected using an average of bias exposures taken at
the beginning and end of the night, adjusted to the overscan region of each
image. The exposures were then divided by internal lamp flatfields, that
were averaged and normalized to a value of unity for each slitlet individually, 
in order to correct for wavelength response.
The flatfielded 2--D images were then corrected for the instrument x and y
distortions by constructing a distortion map from the flatfield and
comparison lamp exposures that was then applied to the science exposures. 
The wavelength calibration and sky subtraction was carried out by the
EXPECTOR program. The wavelength solution was obtained from 25--30 Hg, Ar, Ne,
and Kr lines in the comparison lamp spectra taken before and after the series 
of science exposures and shifted versus the night sky lines on the sciences
exposures. The wavelength solution fitted by a third order Legendre
polynomial, had a typical 1$\sigma$ rms below 0.2 \AA\ ($\simeq 10$ km$\cdot$
s$^{-1}$) in the 
region redward of 4800 \AA\ where it could be well--anchored to night sky
lines. Below 4500 \AA\ the wavelength solution was less well--defined and
could deviate at the blue end by up to 1 \AA. 
The sky subtraction blueward of 5500 \AA, where only a few weak sky
lines are present, worked very well and between
5500\AA\ and 6100\AA\ only the expected photon noise from
the brightest Oxygen and Sodium night sky lines remained after the
subtraction. 
The individual wavelength--calibrated spectra were then extracted under 
IRAF and averaged to produce high signal to noise spectra.

The flux calibration was done using long slit data of the flux standard
Feige 25, observed before the target exposures on the same nights as the
targets.
The response curve seems to vary slightly with the
position of the spectra on the chip, so that velocities and line indices
were measured both on the flux calibrated and flux uncalibrated (hereafter
fluxed and unfluxed) spectra. For all 
objects the
results agreed within small fractions of the errors. Values quoted below for
velocities and line indices are the average of the two methods.

%%%%%%%%%%%%%%%%%%%%%%%%%% VEL %%%%%%%%%%%%%%%%%%%%%%%%%%%%%%%%%%%%%%%%
%%%%%%%%%%%%%%%%%%%%%%%%%%%%%%%%%%%%%%%%%%%%%%%%%%%%%%%%%%%%%%%%%%%%%%%

\section{Radial Velocities of the globular clusters}

\subsection{Radial velocity measurements}

The final spectra were cross-correlated with two velocity template spectra
obtained during the same run (M31 globular clusters 225-280, $v_{helio}=-164$ 
km$\cdot$s$^{-1}$ and 158-213, $v_{helio}=-180$ km$\cdot$s$^{-1}$) using the 
method of Tonry \& Davis (1979)
implemented in the FXCOR package within IRAF. Only 
the spectral region between 4800 \AA\ and 6000 \AA\ was used as this is
where the wavelength solution is best defined. In most cases the
cross--correlation peaks were very well defined and the formal errors returned
by the cross--correlation were on the order of 25 km s$^{-1}$. Combined
with the errors in the wavelength calibration, we estimate the random error
to be on the order of 30 km s$^{-1}$. We checked for any systematic errors
in the reduction of the individual objects by using individual exposures
instead of the combined spectra, and using different comparison lamp spectra
for the wavelength calibration. The dispersion of these individual
measurements was added in quadrature to the formal error. This total error is
given in Table 1, together with the heliocentric velocity, V
magnitude and V-I color taken from the NW field of Kissler-Patig et al.~(1997),
the photographic B$_j$ magnitude and B$_j-$R color from the original list used
by Grillmair et al.~(1994) and, for the five
objects in common, the velocity quoted by Grillmair et al.~(1994). Note
that the objects \#4, \#8, and \#16 turned out to be a star and two
low--redshift early--type galaxies, and will not be discussed in the
following.  Figure 1
shows our velocities versus those of Grillmair et al.~(1994) for the five 
objects in common. The velocities are in good agreement, within 
the errors, and no
systematic shift can be identified in this sample. 
%
%\placefigure{fig1}
%
We plotted the location of our globular clusters with respect to the galaxy
in Fig.~2, where the rings indicate 1 and 5 galactic effective radii 
(taken from Goudfrooij et al.~1994) and the sizes of the symbols reflect
the globular cluster velocities with respect to the mean sample velocity 
(open being approaching, solid being receding). 
Our globular clusters are located between 2 and 7
effective radii of the galaxy and extend from the North to the West of the
galaxy.
%
%\placefigure{fig2}
%

\subsection{Globular cluster velocities}

A histogram of the globular cluster velocities is shown in Fig.~3. The
mean velocity of our sample is 1293 $\pm 71$ km$\cdot$s$^{-1}$, that is 
offset by 150 km$\cdot$s$^{-1}$ from the systemic velocity of NGC 1399 (1447
km$\cdot$s$^{-1}$, de~Vaucouleurs et al.~1991). The mean velocities of
previous samples are 1517 $\pm 91$ km$\cdot$s$^{-1}$  
(Grillmair et al.~1994) and 1353 $\pm 79$ km$\cdot$s$^{-1}$ (or 1309 $\pm 71$
km$\cdot$s$^{-1}$ without an outlier, Minniti et
al.~1998). The latter sample is mainly situated on the opposite side of
the galaxy from
ours (South East of the galaxy), making it unlikely that rotation is the 
cause of our low
mean velocity. This is further supported by the results of Grillmair 
et al.~(1994), who found no evidence for rotation in their more uniformly 
distributed sample.

The velocity dispersion derived from our sample is 302 $\pm 51$ 
km$\cdot$s$^{-1}$.
It is slightly lower than, but consistent within the errors with, the ones 
derived by 
Grillmair et al.~(1994): $388 \pm 54$ km$\cdot$s$^{-1}$, and  Minniti
et al.~(1998): $338 \pm56$ km$\cdot$s$^{-1}$ (or $293 \pm50$ km$\cdot$s$^{-1}$
without an outlier). As noted in the previous studies, the velocity
dispersion is higher than measured for the stellar component
within 1.5 arcmin, suggesting that either the mass--to--light
ratio and/or radial anisotropy changes dramatically between 1.5 and 5
arcmin. 

Figure 4 shows magnitude and color versus velocity. Figure 5 shows 
velocity versus radius. Given the small size of our sample only 
very clear correlations would be discernible. Apparently velocity does not 
strongly correlate with radius, color or magnitude of the globular clusters.
Such correlations will be explored more fully in a forthcoming paper 
which will including all three (Grillmair et al.~1994, Minniti et al.~1998, and this work)
globular clusters samples (Kissler-Patig et al.~1998).

\subsection{Mass estimate and the Mass--to--Light ratio}

\subsubsection{Mass estimate}

The velocity distribution of the globular clusters can be used to 
give an estimate of the mass of 
the parent galaxy. Given the small number of velocities and the poor
spatial coverage, we follow a simple method also applied by Huchra and
Brodie (1987). 
We estimate the mass from the virial theorem on the one hand,
%(e.g.~Zwicky 1957), 
and from the projected mass method on the other hand (Bahcall \&
Tremaine 1981, Heisler, Tremaine \& Bahcall 1985). The virial theorem mass 
for a system of $N$ globular clusters with measured velocities is 
\begin{equation}
M_{\rm VT}=\frac{3 \pi N}{2G} \frac{\Sigma_{i}^{N} V^{2}_{i}}
{\Sigma_{i<j}1/r_{ij}} 
\end{equation}
where $r_{ij}$ is the separation between the $i$th and $j$th cluster, and
$V_i$ is the velocity difference between the $i$th cluster and the mean
system velocity. The projected mass is
\begin{equation}
M_P=\frac{f_p}{G(N-\alpha)}(\Sigma_{i}^{N} V_{i}^{2}r_{i}) 
\end{equation}
where $r_i$ is the separation of the $i$th globular cluster from the centroid. 
The quantity $\alpha$ was chosen to be 1.5 following Heisler et al.~(1985),
as a slight correction to $f_p$, used in a discrete model but determined
analytically from a continuous model.
The quantity $f_p$ depends on the distribution of the orbital eccentricities
for the globular clusters, it ranges from to $64/\pi$ for radial
orbits, over $32/\pi$ for isotropic orbits, to $64/3\pi$ for circular orbits,
in an extended mass. We adopt a projection factor of $32/\pi$, which 
assumes purely isotropic cluster orbits in an extended mass, for
comparison with the simulations of Hernquist \& Bolte (1993).

The results from both methods are: $M_{\rm VT}=2.0(\pm0.9)\cdot10^{12} M_\odot$ 
and $M_P=4.3(\pm1.0)\cdot10^{12} M_\odot$ within a radius of 5 arcmin or 
28 kpc, adopting a distance to NGC 1399 of 19.3 Mpc (Madore et al.~1997).  

The errors were calculated by a standard jackknife procedure (e.g.~Efron \&
Tibshirani 1993). Further, we
have to consider the incomplete spatial coverage of our sample. We
derived the effective radius of the globular cluster system by fitting a 
de~Vaucouleurs profile to the globular cluster surface density data 
of Kissler-Patig et al.~(1997). We obtained an effective radius r$_{\rm
eff(GCS)}=120"\pm20"$. The mean radius of the spectroscopic sample
is $\simeq 1.6$
r$_{\rm eff(GCS)}$. Richestone \& Tremaine (1984) showed that a single
velocity
dispersion observed at 1.6 r$_{\rm eff}$ (approximating our
observations) can vary 
by a factor of $\simeq 1.8$. Thus, 
mass--to--light ratios derived from observations like ours
can vary by a factor $\simeq 3.2$. An additional
error might be present in the projected mass method: Hernquist and Bolte
(1993) simulated a globular cluster distribution around a galaxy and noted
that a sample like ours (roughly 20 objects extending out to 30 kpc,
assuming $f_p=32/\pi$), systematically overestimates the total galaxy mass
by up to 50\%. In addition, the value we adopted for the
projection factor was assumed without knowledge of the real globular
cluster orbits. 

\subsubsection{The Mass--to--Light ratio}

The total luminosity within 5 arcmin is obtained from the light profile of
Bicknell et al.~(1989), correcting for our assumed distance of 19.3 Mpc.
We obtain a total integrated luminosity within 5 arcmin of
$5.5(\pm0.5)\cdot10^{10} L_{\odot \rm B}$ and estimate a mass to light ratio 
within 28 kpc of $M/L_B\simeq36\pm20 M_\odot/L_\odot$ using $M_{\rm VT}$ or 
$M/L_B\simeq76\pm40 M_\odot/L_\odot$ (likely to be an overestimate, see
above)
using $M_P$. 

Grillmair et al.~(1994) derived $M/L_B=79\pm20 M_\odot/L_\odot$, but
assumed a distance to NGC 1399 of 13.2 Mpc, i.e.~their value would
fall around $M/L_B\simeq35M_\odot/L_\odot$ for our assumed distance.
Minniti et al.~(1998) derived $M/L_{\rm B}$ values between 50 and 130, also
assuming a slightly shorter distance. All samples cover comparable 
galactocentric distances.  In summary the values agree when corrected for the
different assumed distances. The main result is that this
mass--to--light ratio is about a factor 10 above that expected for an old stellar
population, leading to the conclusion that, at this distance from the
center ($\simeq 6$ r$_{\rm eff}$), dark matter dominates the potential.  
The power of globular clusters for such studies was nicely illustrated 
by Cohen \& Ryzhov (1997), who computed $M/L$ ratios at various radii from
a large sample of globular cluster velocities in M87.

Finally, we note that our results are in good agreement with the ones
derived from X--ray data. Jones et al.~(1997) report a total mass for NGC
1399 of 4.3 to $8.1\times 10^{12} M_{\odot}$ within 18 arcmin, and a
mass--to--light ratio increasing from $33\pm8 M_\odot/L_\odot$ at 2.6
arcmin to $70\pm22 M_\odot/L_\odot$ at 15.8 arcmin (for an assumed distance
of 24 Mpc).

%%%%%%%%%%%%%%%%%%%%%%%%% ABUN %%%%%%%%%%%%%%%%%%%%%%%%%%%%%%%%%%%%%%%%%%
%%%%%%%%%%%%%%%%%%%%%%%%%%%%%%%%%%%%%%%%%%%%%%%%%%%%%%%%%%%%%%%%%%%%%%%%%

\section{Element abundances}

\subsection{Measuring the indices}

Absorption line indices were measured following the procedure described
in Brodie \& Huchra (1990) who used the Lick/IDS system bandpasses defined
in Burstein et al.~(1984). Mean heights were defined in each of the
pseudocontinuum regions and a straight line was drawn through their midpoints.
The difference in flux between this line and the observed
spectrum in the index bandpass determines the index.

In the following paragraph we discuss some details of the line measurements 
that can slightly affect the index values.
We note that the Lick/IDS bandpasses as defined in Burstein et al.~(1984)
were recently fine--tuned and the system was extended (see Gonzales 1993 and 
Trager 1997).
Unless otherwise noted, we used the Burstein et al.~(1984) definitions
in order to be compatible with the metallicity calibrations of Brodie \& Huchra
(1990). Further, unless otherwise noted, we did not artificially degrade our 
resolution from 5.6\AA\ to the 8.5-9\AA\ of the Lick/IDS system, again to stay 
consistent with Brodie \& Huchra (1990) whose most relevant data for our
comparison were uncorrected for resolution effects and had 5\AA\ resolution 
(the rest of the Brodie \& Huchra data spanned the range from
5\AA\ to 12\AA\ resolution).  We tested the effects of
resolution by measuring indices from our original spectra and from our
spectra degraded, with a Gaussian filter, to 9\AA\ resolution. As
already noticed by Gonzales (1993), the broader indices (e.g.~Mg2) are
barely affected, while narrower indices (Fe, H$\beta$, ...) are
systematically lower (by about $0.006\pm0.002$ mag in our case) when
measured from the lower dispersion spectra. This amounts only to $\simeq 30\%$
of our photon noise error, but it is systematic. 
We further tested our index measurement routine versus that used by the
Lick/IDS group in order to check the determination of the continuum and the
treatment of the bandpass end--points. 
For this purpose Worthey provides a set of high signal--to--noise templates
and a list of index bandpasses measured by the Lick/IDS software
(http://www.astro.lsa.umich.edu/users/worthey/) on which one can test one's
own program. We found a perfect agreement between our software and Worthey's
(the original used to measure the indices in the Lick/IDS system). 
Finally, we checked for
differences using the new and old Lick/IDS bandpasses (typically shifted by
0.5-2.6\AA\ from the original definitions). No systematic effect could be
found for most indices, the variations depending on the photon noise in 
individual resolution elements of the spectra. However we 
found the index values for Fe5335 (whose 
definition was shifted by 2.6\AA\ on the blue side) to be systematically
lower (by $\simeq 0.01$ mag) when using the new definition. 
We conclude that care should be taken when inter--comparing measurements of
various groups. An exact comparison of the values quoted in 
Table 2 with the latest Lick/IDS system is only possible after correcting 
for the systematic offsets.
Finally we point out that all indices in Table 2 are given in magnitudes,
while the Lick/IDS system usually quotes the atomic indices in \AA.
Further, in the same table, the values for NGC 1399 taken from Huchra et
al.~(1996) were measured on the spectrum without deconvolving it with the 
velocity 
dispersion of the galaxy ($\simeq 350$ km s$^{-1}$). The effect is very
similar to degrading the resolution of the spectrum by almost a factor two, 
and the measured line indices are therefore systematically lower than 
Trager's (1997). In the following we will use the data for NGC 1399 from
Trager (1997) when comparing the galaxy with the globular clusters.  

Representative fluxed spectra of a blue, a red and an extremely red globular 
cluster are shown in Fig.~6. The spectra displayed were smoothed with a 3 pixel 
average filter. We measured all our indices on the fluxed and unfluxed
spectra. The duplication test was carried out because of the 
uncertainties of flux calibrating multi--slit data with flux standards
taken in long slit mode.
Flux calibrating is a multiplicative procedure and should
not, therefore, affect the line indices. Only broad indices could be affected
if the slope of the continuum dramatically changes from the unfluxed to the
fluxed spectrum. 
We compared the indices measured on the flux and unfluxed spectra and found
the differences to be negligible. All values given
in Table 2 are averaged from the results from the fluxed and unfluxed spectra.
The errors were estimated from the photon noise in the bandpasses (see
Brodie \& Huchra 1990) and from the error in the wavelength calibration for
indices blueward of 4500 \AA\ (only G band). All spectra had comparable
signal--to--noise and the errors were found to be similar for all objects at 
a given index.

\subsection{Element abundances of the globular clusters}

In Fig.~7 we show the relations between Mg2 and various other indices,
together with the range covered by the Milky Way and M31 globular clusters
(shown as shaded areas). 

The first point to note is that the globular clusters in NGC 1399 (except
for two clusters that we will discuss further below) cover the {\it
entire} range spanned by Milky Way and M31 globular clusters. They do not show 
anomalies in the metal--tracing indices such as Mg and Fe, or in the 
more age--sensitive index, H$\beta$. Apparently NGC 1399 mostly hosts globular 
clusters with ages and metallicities similar to those found in the spirals of
the Local Group.

In Fig.~8 and 9 we plot Mg2 versus the equivalent width of Fe5270, 
Fe5335 and H$\beta$ in order to compare them to population synthesis
models, to the value of the host galaxy NGC 1399, and to similar plots for
elliptical galaxies (e.g~Worthey et al.~1992). The indices of Table 2 are
shown as solid dots. For clusters \#2 and \#14 we also included the
indices measured in the new Lick/IDS system (new
bandpass definitions, resolution degraded to 9 \AA) as open circles.
The values for NGC 1399 itself are taken from Trager (1997) and shown as 
triangles. For completeness we show in Table 2 also the line indices
derived by Huchra et al.~(1996) for NGC 1399. Note that these latter did
not deconvolve the galaxy spectrum with the velocity dispersion of the
galaxy before measuring the line indices. This leads to systematically
lower values of Huchra et al.~(1996) and prevents a direct comparison with
Trager's or the globular cluster measurements.
In Fig.~8 the population synthesis
models of Fritze-v.~Alvensleben \& Burkert (1995) are shown as long dashed 
lines for 8 and 16 Gyr year old populations, with metallicity varying between
Z=0.001 and Z=0.04. Worthey's (1994) models are plotted as short dashed lines 
for 8 and 17
Gyr and metallicity varying between $[$Fe/H$]=-2.0$ and $0.5$ dex.
The range covered by the Mg--rich elliptical galaxies of Worthey
et al.~(1992) is shown as a hatched area.
The same symbols are used in Fig.~9, except that we show the tracks for 
16, 8, and 3 Gyr for models of Fritze-v.~Alvensleben \& Burkert (1995), and
for 17, 8, 3, and 1.5 Gyr in the case of Worthey's (1994) models.

\subsubsection{Mg2 versus Fe}

The observed relations between Fe and Mg2  are well--reproduced by the
population synthesis models of Fritze-v.~Alvensleben \& Burkert (1995)
and Worthey
(1994). Given the small age dependence of these indices and our errors, we
cannot discriminate age differences from this plot (Fig.~8). Worthey et
al.~(1992) report a Mg versus Fe enhancement for large ellipticals.  
Unfortunately we have only two globular clusters with Mg high enough to
compare them with the affected ellipticals. Given our errors on Fe, these
two globular clusters are compatible with the models and also fall in
the range covered by the Mg--rich ellipticals. The other element observed
to be unusually high in bright ellipticals is NaD. Indeed the two metal--rich 
clusters, especially \#14, have high NaD abundances that could 
hint to abnormal high values. In summary, both globular cluster \#2 and \#14 have
values close to the ones measured for the stars in NGC 1399.

More data for Mg rich
globular clusters are clearly needed to investigate this interesting point:
Since globular clusters can be as rich in Mg  as large ellipticals, their
study might constrain the scenarios responsible for the Mg versus Fe
enhancement in large ellipticals. 
Worthey et al.~(1992) summarize the three mechanisms that can
be responsible for an enhanced $[$Mg/Fe$]$ ratio in large ellipticals versus
normal ratios in smaller ellipticals: {\it i)} Different star formation 
timescale i.e.~no time for SNe Ia to dilute the abundance ratios set by the 
earlier SNe II; {\it ii)} Variable/flatter IMF, e.g.~in mergers, to produce 
more SNe II versus SNe Ia; {\it iii)} 
Selective loss mechanisms retaining Mg and expelling Fe in giant
ellipticals (this last scenario is rather unlikely as they note). 
While selecting between these scenarios is difficult on the basis of the
properties of the starlight alone, globular clusters might help to settle 
the issue.  If, for example, the Mg enrichment is a
result of a flatter IMF for star formation in mergers, globular clusters formed
in mergers will show the Mg over--enrichment too. If, however, the high
$[$Mg/Fe$]$ values are the result of different star formation timescales
among ellipticals, and the Mg--rich globular clusters are young 
(a possibility discussed below), e.g.~formed in a later merger, then they will 
probably not show the higher $[$Mg/Fe$]$ ratios.

\subsubsection{Mg2 versus H$\beta$}

The relation between Mg2 and H$\beta$ is interesting in many respects.
First it shows the significant differences in the predictions from various 
models for
the value of H$\beta$ at a given age and metallicity. A discussion of the
causes of these differences is out of the scope of this paper. We concentrate 
on the points in common to all models. These are, on the one hand that our 
globular clusters (except for the two outliers) could span a large range of age
according to the models. Our error in H$\beta$ does not allow an age
determination better than within a factor two, but our
measurements are also consistent with a single age, within the errors. 
And we note 
that our values lie in the range spanned by the Milky Way globular
clusters (see Fig.~7), that span a small (few Gyr) range in age. The 
NGC 1399 globular clusters could, therefore, all
have similar ages and be as old as the globular clusters in the Milky Way.

On the other hand, we note that the H$\beta$ values for our two metal rich
clusters cannot be reproduced by any model, no matter what age or
metallicity is assumed. Large values for H$\beta$ were already reported
(at lower Mg) in some M31 globular clusters (Burstein et al.~1984,
Brodie \& Huchra 1991).
To check alternative explanations to age for strong H$\beta$ values,
Burstein et al.~(1984) computed a semi--empirical model to estimate the
effects on H$\beta$ and Mg when taking a cluster like 47 Tuc or M71 as
starting point (both having stubby, red horizontal branches) and changing
the horizontal branch to be blue like that of M5 (so as to yield
maximum H$\beta$ strength). Their result shows that H$\beta$ can be raised
roughly from 1.5 \AA\ to 2.5 \AA,
without significantly affecting the Mg abundance.
Quantitatively, this alone could explain the position of globular cluster
\#2 and \#14 in Fig.~9.
This is very interesting in the light of the extended blue horizontal
branches found recently by Rich et al.~(1997) in metal rich Milky Way
globular clusters. Blue horizontal branches are unexpected (and were, until
recently, not observed) at typical bulge metallicities (e.g.~Fusi-Pecci et
al.~1992, 1993). This is probably the
reason why the contribution of a blue horizontal branch is not taken into 
account in any of the models at high metallicities even though it can
contribute to 
stronger H$\beta$ values. To date, in the models of metal--rich stellar
populations, most  of the contribution to H$\beta$ and the blue light came
from stars from the turn--off region,  so that variations in
color or H$\beta$ had to be explained by the position of the turn--off
point, i.e.~by variations in age. This necessarily created a
strong age--metallicity degeneracy. New models which include the possible
contribution of blue horizontal branches are clearly needed to cover the whole
of the possible abundance range  at given age and metallicity for globular clusters.
More data for globular clusters at high Mg and H$\beta$ might help in the
future to fine--tune the populations synthesis models.

\subsubsection{The most metal rich clusters}

Our two clusters \#2 and \#14 are outliers in many respects and bring
insight into some of the properties of very metal rich globular clusters.
Unfortunately we do not have a V$-$I color for our cluster \#2 and
the B$_j-$R color is too uncertain to rely on (e.g.~compare colors for \#10 
and \#20). However, \#14
is the reddest cluster in the sample and falls in the very red tail of the
color distribution (Kissler-Patig et al.~1997), hinting that
such objects are rare. Further these two objects show that
globular clusters can have solar abundances (see Table 3) but they are
not much in excess of solar (except perhaps for their Mg abundances). This is
contrary to results from broad--band photometric studies (Johnson \&
Washington photometry) in several galaxies (e.g.~Secker et al.~1995 for the
most extreme case) where
globular clusters were reported as having metallicities up to four times
solar. While the abundances of the rest of the clusters in NGC 1399 are
similar to abundances found in the Milky Way and M31, the abundances of 
clusters \#2 and \#14 are clearly anomalous.  

We note that the measurement of the H$\beta$ index is controlled in a 
negative sense by Mg
(Tripicco \& Bell 1995). Thus, in our particular case,
H$\beta$ could even be slightly underestimated. 
In order to further check the strength of the Hydrogen lines, we measured
H$\gamma$ for clusters \#2 and \#14, using the bandpasses defined by 
Worthey \& Ottaviani (1997) extending the Lick/IDS system (H$\gamma _A$ for
young stellar populations, H$\gamma _F$ for old stellar populations). For this 
purpose we degraded the resolution of our spectra to 9 \AA\ and re--determined
the wavelength calibration around the H$\gamma$ line, before measuring the
indices. In both cases we were close to the blue limit of our spectral
range in a region showing a factor 10 less signal than in the region
between 4800\AA\ and 6000\AA. We obtained values of H$\gamma _A =
-4.0\pm 5.0$ \AA\ and $-3.7\pm 2.0$ \AA\ and  H$\gamma _F = 2.1\pm 2.2$ \AA\
and $-0.5\pm 1.2$ \AA\ for clusters \#2 and \#14 respectively. Despite the
large errors, all the values are unexpectedly high at the high Mg values and
are incompatible with high ages ($>8$ Gyr) according to Worthey \& Ottaviani
(1997).
Further, the blue continuum bandpass for H$\gamma$ exactly covers the G--band, 
which is strong in both clusters, so that our H$\gamma$
indices are probably underestimates when compared to young stellar
systems.

It remains unclear how to interpret these strong Balmer lines, and in particular
the high H$\beta$ values. As discussed above, they could be explained by 
the presence of blue horizontal branches (which would also account for the 
location of the clusters in Fig.~9) and/or the strong Balmer lines could be,
at least partly, associated with a younger age but new models would have to
be computed for a reliable comparison. A young age cannot, by
itself, explain the location of the clusters in Fig.~9. For ages
as young as 1 to 5 Gyr, one would expect diluted metal lines and blue 
broad--band colors, which are definitively not seen in these clusters. 
However, a younger age would be consistent with the very high 
metallicities of these clusters (see next section) in a picture
where they formed later than the other clusters from highly enriched gas. 

Their absolute luminosities provide no further clues.
The difference between an 15 Gyr and 2 Gyr old cluster (same mass and same
metallicity) would be on the order of 2 magnitudes. These clusters are not
exceptionally luminous within our sample.
They have absolute V luminosities around $-10.3$ given our assumed distance.
They could be old and have masses as high as Omega Centauri (V$=-10.16$, the 
most massive globular cluster in the Milky Way) or be younger and have 
lower, more ``normal'', masses.

Finally, we note that three of our other globular clusters show Mg values
at the upper limit of the range spanned by the Milky Way and M31. Two of these
clusters also exhibit slightly stronger H$\beta$ values, than expected for
old clusters (although not significantly when considering the errors). This might be yet
another sub--population. Clearly, larger samples of globular clusters with
well--defined abundances will be valuable in exploring in more detail
globular cluster systems with complex formation histories.

\subsection{Deriving metallicities}

We derived $[$Fe/H$]$ values for the globular clusters using our indices  
and following the method of Brodie \& Huchra (1990). We applied their 
index--$[$Fe/H$]$ 
relations to various elements and listed the results in Table 3. We include
values for the G--band, CN, and the Ca H$+$K lines for the few spectra extending
far enough to the blue, as well as NaD where the night sky lines could be
well enough subtracted. Brodie \& Huchra (1990) derived the relations
from Milky Way and M31 clusters, i.e.~the relations
do not extend to the high metallicities seen for some of the globular
clusters in NGC 1399. In particular, we found the relations between Mg2 
and $[$Fe/H$]$, as well as between MgH and $[$Fe/H$]$ to be
badly approximated by a linear extrapolation at higher metallicities 
(Mg2$>0.180$) when compared to population synthesis models (see Fig.~10). 
For these two relations we applied a correction to the five and two most
metal--rich globular clusters respectively by allocating them the
metallicity predicted by the 17 Gyr models at their Mg abundance.  
In addition to the $[$Fe/H$]$ values from individual indices
we computed a mean $[$Fe/H$]$ value from Fe5270 and Mg2 (shown in column
10 of Table 3) and a weighted mean $[$Fe/H$]$ from all available indices (column 11).
The weights were chosen to reflect the quality of the
calibrating relation on the one hand, and the quality of the measured index
on the other hand. We assigned weights of 1 to Mg2 and Fe5270, and 0.2
to all other available indices. We estimated
the errors from the errors on the indices. The error on the final weighted
mean metallicity is estimated to be 0.20 dex for most clusters, slightly
higher ($\simeq 0.30$ dex) for the clusters with high Mg2. 

Brodie \& Huchra (1990) calibrated their index--metallicity relations using
the metallicities from Zinn \& West (1984), which are mainly based on the 
Ca II K line. The Zinn \& West scale
was recently claimed to be slightly non--linear compared to  the total
metal abundance scale (Carreta \& Gratton 1997). We emphasize that our
results are necessarily tied, via the Brodie \& Huchra (1990) calibration, 
to the Zinn \& West (1984) results.

In Fig.~11 we plotted $[$Fe/H$]$ values derived from Fe5270 and Mg2 versus
each other and versus the weighted mean $[$Fe/H$]$ values. We also show a
histogram over the weighted mean metallicity values for our clusters.
The $[$Fe/H$]$ values derived from Mg for high metallicities lie
somewhat above the values derived from Fe5270. As already mentioned, this 
is mostly
due to the fact that, for these high Mg2 values, the relation defined by
Brodie \& Huchra (1991) is no longer valid, and our correction crude. 
It could also be due to a
slight over--enrichment of Mg versus Fe, but our data do not allow us
to make any strong statement in this regard (see last section). In Fig.~11
(lower left panel), three additional clusters seem to have deviating Mg/Fe 
ratios, however this trend is not confirmed by their Fe5335 values in Fig.~8 
(lower panel) where these clusters show normal Fe5335 versus Mg2
abundances. For
metallicities below -0.3 dex the agreement between the $[$Fe/H$]$ value
derived from Mg2 and Fe5270 is good. 

The histogram over the weighted mean $[$Fe/H$]$ values shows that our
sample includes globular clusters with metallicities ranging from
typical values for Milky Way halo globular clusters to slightly above
solar but there are no objects with significantly super--solar abundances. 
Very metal poor clusters ($[$Fe/H$]<-1.6$) appear to be missing, but recall 
that  the blue part of the globular color distribution in NGC 1399 is not 
well--sampled. The mean metallicity for our sample is
$[$Fe/H$]=-0.83\pm0.13$ dex (the error is the standard error of the mean).
The prediction from the galaxy luminosity -- mean globular cluster
metallicity relation of Brodie \& Huchra (1991) is $[$Fe/H$]=-1.14\pm0.12$,
or $[$Fe/H$]=-0.91\pm0.16$ (when using the relation defined from
ellipticals only),
adopting, for NGC 1399, B$_{\rm T}=10.55$ mag (de Vaucouleurs et al.~1991) and 
a distance modulus of 31.43 (Madore et al.~1997).

Because of the size of our spectroscopic sample, we have to return to 
broad--band colors to comment further on the metallicity distribution in 
the globular cluster system of NGC 1399.

%%%%%%%%%%%%%%%%%%%%%%%%%%% V-I vs indices %%%%%%%%%%%%%%%%%%%%%%%%%%%%%%
%%%%%%%%%%%%%%%%%%%%%%%%%%%%%%%%%%%%%%%%%%%%%%%%%%%%%%%%%%%%%%%%%%%%%%%%%

\section{Broad--band colors versus indices and metallicity}

In this section we compare the line indices with the broad--band
colors of our globular clusters. We present the indices versus our V$-$I
colors from Kissler-Patig et al.~(1997), for which the typical error 
on the color is 0.035 mag at these magnitudes.
Being of photographic origin, the
uncertainties in existing B$_j-$R colors (from Grillmair 1992)
listed in Table 1 are too large (of the order of 0.3 mag) for our purposes.

%\placefigure{}

In Fig.~12 we show our V$-$I colors versus Mg2, Mgb, $<$Fe$>$ (defined as the
mean of Fe5270 and Fe5335, as introduced by Burstein et al.~1984), and H$\beta$.
We note that the broad--band color correlates well with the metal
indices. This might have been expected since neither Mg or Fe are very age
sensitive (e.g.~Worthey 1994), so that age--metallicity degeneracy is
not a factor here. However, we note that V$-$I also correlates well (inversely)
with H$\beta$. As in Fig.~9, the V$-$I colors scatter by less than their
typical errors, and are therefore consistent with a single age
for these clusters. But, as already noted above, the limiting factor in
deriving ages are our errors in H$\beta$ which span a factor of two in age 
according to population synthesis
models (e.g.~Fritze-v.~Alvensleben \& Burkert 1995, Worthey 1994).

Formal linear fits return the following relations:\\
Mg2 $=-0.46(\pm 0.09) + 0.57(\pm 0.08)$V$-$I \\
$<$Fe$>$ $=-0.08(\pm 0.02) + 0.12(\pm 0.02)$V$-$I \\
H$\beta = 0.17(\pm 0.03) - 0.07(\pm 0.03)$V$-$I \\
%\placefigure{fig V-I vs Fe/H}

In Fig.~13 we plot our V$-$I colors versus the weighted mean [Fe/H] (see
above section and Table 3) as filled triangles, together with the values for all
Milky Way globular clusters that have a reddening of less than E(B$-$V)=0.2
(open circles).  The Milky Way values were taken from the McMaster database 
(Harris 1996) and de--reddened
according to Rieke \& Lebofsky (1985). For our cluster \#14 we plotted the
metallicity derived from Fe5270, as an open triangle. For the first time, 
insight can be gained into the relationship between globular cluster 
metallicities and colors for the reddest clusters (V$-$I $>1.1$ mag).

We note that the relation may be slightly non--linear in the sense that the
slope gets flatter to redder colors, as predicted by the population
synthesis models. While the V$-$I color reflects well metallicity,
as shown above, it could still be affected by age within our errors. 
We stress again that the ages of our
globular clusters are not well defined, but they are compatible
with the ages of Milky Way globular clusters, as inferred from all the
measured element abundances. The most metal--rich globular clusters may have 
formed later than the rest. If they are, for
example, half as old as the metal poor ones,
their V$-$I color is shifted to the blue by 0.1 magnitude. That is, if we
corrected the colors of our most metal rich globular clusters for age,
we would get an even flatter slope. However since the ages are  uncertain,
we will not apply any corrections for age to the colors in what follows.

Given the restricted number of data points for
colors above V$-$I=1.2, we attempted only a linear fit to the sample
(including the selected Milky Way globular clusters). The
returned relation between metallicity and V$-$I is:\\

$[$Fe/H$]$ $=-4.50(\pm 0.30) + 3.27(\pm 0.32)$V$-$I\\

The slope of this relation is almost twice as flat as the ones previously derived 
by Couture et al.~(1990) and Kissler-Patig et
al.~(1997) from the Milky Way globular clusters alone.
A non--linear fit will lead to an even more dramatic result. For red colors, 
metallicities derived from V$-$I, and most probably from other broad--band 
colors too, have been overestimated. Since most globular cluster systems in 
early--type galaxies have red median colors (V$-$I $>1.1$), most mean 
metallicities have probably been overestimated in the past.

Kissler-Patig et al.~(1997) have shown that the color distribution of NGC
1399 has two ``peaks" at V$-$I $=0.99$ mag and V$-$I $=1.18$ mag (confirming
the multi--modal distribution found with Washington photometry by
Ostrov et al.~1993, also seen by Forbes et al.~1997). Thus, if these two peaks are associated with two different sub--populations, these two
populations would have mean metallicities
around [Fe/H]$=-1.3$ dex and [Fe/H]$=-0.6$ dex, very similar to the means of
the halo and disk/bulge populations of the Milky Way. 
The same applies to globular clusters in M87 which
peak at V$-$I$=0.92$ and V$-$I$=1.23$ (Elson \& Santiago 1996). This
corresponds to [Fe/H]$=-1.5$ dex and [Fe/H]$=-0.5$ dex, similar to NGC 1399 and
again similar to the Milky Way.  Further, the small
ellipticals NGC 1374, NGC 1379, NGC 1387, NGC 1427 which appear uni--modal
(Kissler-Patig et al.~1997) would have mean metallicities around 
[Fe/H]$=-0.9$, $-0.7$, $-0.6$,
$-1.1$ dex respectively. That is, their globular clusters do not have 
solar or super--solar mean abundances as previously thought. 
Moreover, the spread in metallicity between the different galaxies
is reduced to 0.5 dex instead of 1 dex, leading to a more homogeneous picture
and smaller discrepancies between metal--rich globular clusters in spirals and
ellipticals.

Finally we note that very metal--rich globular clusters such as our objects
\#2 and \#14, with V$-$I colors $\geq 1.35$ make up 
only a small fraction (2--5\%) of the total globular cluster system, as
derived from the color distribution of Kissler-Patig et al.~(1997). If
these objects are significantly younger, i.e.~brighter, they might even be
over--represented in the magnitude limited sample of Kissler-Patig et
al.~(1997) and represent an even smaller fraction of the total globular
cluster system.

We conclude from the above that V$-$I, and thus probably the other broad--band
colors too, trace metallicity relatively well in globular cluster systems.
Drawing a tentative conclusion from this sample, the V$-$I relation 
can be used if most objects are old and do not show a significant age
spread, as it seems the case for our sample that scatters around an
isochrone within the measurement errors (see section 4). Generally, this
will be the case for any globular cluster system that formed at high redshift 
with no significant fraction of clusters that formed since $z \simeq 1$.
Further, given the spread in [Fe/H] at a given color, metallicities from V$-$I 
will be accurate only to 0.5 dex for individual objects. However, mean
metallicities for entire globular cluster systems can probably be derived
with an accuracy of 0.3 dex from V$-$I.
This will, for example, allow corrections for metallicity effects 
to distances derived from the globular cluster luminosity function, 
when the mean color of the
globular cluster system is known, as proposed by Ashman, Conti \& Zepf
(1995).

%%%%%%%%%%%%%%%%%%%%%%%%%% DISCU  %%%%%%%%%%%%%%%%%%%%%%%%%%%%%%%%%%%%%%
%%%%%%%%%%%%%%%%%%%%%%%%%%%%%%%%%%%%%%%%%%%%%%%%%%%%%%%%%%%%%%%%%%%%%%%

\section{Discussion}

Most globular clusters in
NGC 1399 have very similar Mg, Fe and H$\beta$ line indices to the Milky Way 
and M31 globular clusters and span the full range observed in these
galaxies. The metal--poor clusters have metallicities corresponding to the mean
metallicity of Milky Way halo clusters, the metal--rich clusters have
similar metallicities to Milky Way bulge/disk clusters. Their Mg and Fe and especially
H$\beta$ indices indicate that these clusters are probably as old as the Milky
Way globular clusters, although deriving accurate ages is difficult given
the observational errors and model uncertainties.

Two clusters clearly stand out in their abundances. They have Mg and Fe
(and Na) indices significantly higher than observed in any globular 
cluster in the Milky Way or
M31, and comparable to the integrated starlight of giant ellipticals.
Their metallicities are estimated to be around, or even slightly above,
solar. Whether or not these clusters exhibit a Mg versus Fe
over--abundance, as seen in the integrated starlight of giant ellipticals, 
remains an interesting open question.
Moreover, these two clusters show abnormally high H$\beta$ (and H$\gamma$)
values that cannot
be matched by population synthesis models for any age and metallicity,
making any age determination very uncertain and calling for an explanation. 
Blue horizontal branches in these metal--rich clusters would qualitatively
be a solution.
The formation of these globular clusters several gigayears
after the more metal--poor globular clusters could also be part of the  
explanation but needs revisited models including blue horizontal branches
at high metallicity for comparison.
This would confirm the formation of at least some globular clusters
in a later phase as suggested for globular clusters forming in mergers
(Schweizer 1987, Ashman \& Zepf 1992).
From the globular cluster
color distribution, we estimate that these peculiar clusters do not make 
up more than 5\% of the total population, hinting to mergers
involving little dissipation.

The overall picture of the system seems to hint at the presence of
various sub--populations
similar to the ones seen in the Milky Way, as well as a small fraction of
globular clusters that formed later from highly enriched gas. That is,
processes like the ones that produced the Milky Way halo and disk/bulge
population can be responsible for the formation of the vast majority of the
globular clusters in NGC 1399, only a small percentage of the total number
of globular clusters need to have formed later
from solar metallicity gas. Fritze-v.~Alvensleben \& Gerhard (1994) showed
that, at the end of a starburst induced by a merger
event, the metallicity can reach solar, independently of the age of the event
(under the assumption that no gas is lost from, or accreted by, the merging
galaxy pair). This would allow the formation of the extremely metal--rich
clusters in a merger as soon as 3 Gyr after the formation of the progenitor
galaxies if globular clusters formed at the end of the starburst.
These few new globular clusters would be added to the globular
clusters already present in the progenitors (the majority).

Translated into implications for the formation of NGC 1399 and its
over--abundant globular cluster system: The galaxy and its globular
clusters are likely to have formed at an early time. The blue population formed 
from material as metal poor as the Galactic halo, despite the fact that it
existed in a high
density environment like the center of a cluster of galaxies, and it
presumably formed at a similar
early epoch. The relatively low mean metallicity (compared to previous
estimates, based on broad--band colors) of the ``red'' population 
(peaking at V$-$I=1.18) of $[$Fe/H$]\simeq -0.6$ dex, implies formation 
before the gas could be enriched to solar abundances. Whether these globular
clusters formed in early merger events or during the collapse of the ``bulge'' 
(as might have been the case in the Milky Way, e.g.~Minniti 1995)
cannot be distinguished from our data. However, judging by the globular cluster color
distribution and the metallicity of the very red objects, only a very small
fraction of the globular clusters clearly formed later from solar metallicity
gas, i.e.~later mergers are unlikely to be the cause for the over--abundant
globular cluster system of NGC 1399. To explain the abnormally high number 
of globular clusters and specific frequency of NGC 1399, alternative models
are needed (see e.g.~Forbes, Brodie \& Grillmair 1997, Blakeslee et
al.~1997), and the role of mechanisms like stripping from neighboring 
galaxies (Muzzio 1987), or accretion of dwarf galaxies (Hilker et al.~1997)
must be better understood.

%%%%%%%%%%%%%%%%%%%%%%%%%% CONC  %%%%%%%%%%%%%%%%%%%%%%%%%%%%%%%%%%%%%%
%%%%%%%%%%%%%%%%%%%%%%%%%%%%%%%%%%%%%%%%%%%%%%%%%%%%%%%%%%%%%%%%%%%%%%%

\section{Summary and Conclusions}

We obtained moderate resolution spectra for 18 globular clusters in the giant
elliptical cD galaxy NGC 1399 in Fornax. From the derived velocities we
calculated a galaxy mass of 1 to $5 \times 10^{12} M_\odot$, leading to a mass
to light ratio of $M/L_{\rm B}=36\pm20$ or $76 \pm 40 M_\odot/L_\odot$ 
(depending on the estimator) within
5 arcmin (6.7 r$_{\rm eff}$), for an assumed distance of 19.3 Mpc. This
would imply a dark matter dominated potential at this radius. No
correlation between magnitude, color or position with velocity could be
found in our small sample.

The element abundances of most globular clusters in NGC 1399 do not differ
from the ones observed in the Milky Way and M31. No different processes and
formation time than the ones assumed for the formation of the Milky Way halo 
and disk/bulge globular clusters
are needed to explain the majority of globular clusters in NGC 1399.
However, two of the globular clusters in our sample clearly stand out 
in the strength of  their 
line indices and show abundances similar to the starlight of giant ellipticals.
They must have
formed from solar metallicity gas in a different formation process,
e.g.~merger event, although not necessarily much ($>3$ Gyr) later than the
others.
Further, these very metal rich clusters show unexplained high H$\beta$ (and
H$\gamma$) abundances, incompatible with any age--metallicity combination
of existing models. These abundances can, however, be explained by 
blue horizontal branches. Judged from the color distribution, these clusters 
represent less than 5\% of the total number of globular clusters.
This hint at late mergers not being the cause of the over--abundant
globular cluster system around NGC 1399. 

The age--metallicity degeneracy of broad--band colors, as predicted by
the models, is presumably artificially strengthen by not taking into
account a possible contribution to the blue light from the horizontal
branch at high metallicities.  
Broad--band colors turned out to be very good metallicity tracers in the
globular cluster system. 
While for individual cases the metallicity of a
globular cluster can only be derived to an accuracy of 0.5 dex from its V$-$I color, the
mean metallicity of a globular cluster system can be determined to an
accuracy of about 0.3 dex. However, we stress that the relation derived here has
a slope about twice as shallow as the ones previously extrapolated from the
Milky Way system, making globular cluster metallicity differences between 
galaxies less significant and bringing estimates for red globular 
clusters back from super--solar to solar metallicities.

These conclusions were drawn from a sample of only 18
globular clusters. The picture will certainly improve as more spectroscopic 
studies from 10--m class telescopes appear in the future.

In a recently submitted paper Cohen, Blakeslee \& Ryzhov (1998) present a
similar study in M87. They obtained Keck/LRIS spectra for a 
larger sample of globular clusters and find a similar metallicity range as 
for the clusters in NGC 1399,
ranging from metal--poor to solar metallicities, and an old median age for
their globular cluster sample. 
Their spectroscopy strengthen the point that globular clusters in large
ellipticals have very similar line indices to the globular clusters in the
Milky Way and M31. 

\acknowledgments

We would like to thank the staff at the Keck Observatories, as well as the
entire team of people, led by J.B. Oke and J.G. Cohen, responsible for the
Low Resolution Imaging Spectrograph, for making the observations possible.
We would also like to thank A. Phillips for his help to run his 
slitmask preparation and alignment software. 

Further, we are thankful for interesting discussions and comments from Mike 
Bolte and Sandy Faber. Many thanks also to Uta Fritze-v.~Alvenlesben
for the electronic version of her new models, as well as useful remarks.
Thanks to Dan Kelson for providing his program EXPECTOR prior to public 
release, and to Luc Simard for his help in using it. Thanks to Scott
Trager and Sandy Faber for making their data on NGC 1399
available prior to publication. We thank the referee J.~Secker for some useful
comments.

The research was partly supported by the 
faculty research funds of the University of California at Santa Cruz.

%\appendix

%%%%%%%%%%%%%%%%%%%%%%%%%%%%%%%  REF  %%%%%%%%%%%%%%%%%%%%%%%%%%%%%%%%%%

\clearpage

%%%%%%%%%%%%%%%%%%%%%%%%%%%%%%%%%%%%%%%%%%%%%%%%%%%%%%%%%%%%%%%%%%%%%%%%%
\clearpage

\begin{deluxetable}{c cc rc rc r r}
%\tablewidth{0pc}
\tablenum{1}
\tablecaption{List of our candidate globular clusters around NGC 1399}
\tablehead{
\colhead{ID} & \colhead{RA(1950)} & \colhead{DEC(1950)} &  \colhead{V}(a) &
\colhead{V$-$I}(a) & \colhead{$B_j$}(b) & 
\colhead{$B_j-R$}(b) & \colhead{$v_{\rm helio}$} & \colhead{$v_{G+94}$(c)} \\
& & & $\pm 0.02$ & $\pm 0.035$ & $\pm 0.2$ & $\pm 0.3$ & & 
}
\startdata
1 & 3 36 13.8& -35 39 24.8 & \nodata &\nodata &21.8 & \nodata  &  732 $\pm32$ &\nl
2 & 3 36 14.2& -35 38 51.2 & \nodata &\nodata &22.4 &1.19& 1094 $\pm34$ &\nl
3 & 3 36 09.4& -35 37 32.4 & \nodata &\nodata &22.3 &1.02& 1571 $\pm31$&\nl
4 & 3 36 12.0& -35 37 44.3 & \nodata &\nodata &21.9 &1.20&  279 $\pm72$ &\nl
5 & 3 36 13.2& -35 37 37.8 & \nodata &\nodata &21.8 &1.17& 1775 $\pm66$ &\nl
6 & 3 36 17.8& -35 37 50.2 & \nodata &\nodata &22.3 & \nodata  & 1386 $\pm31$&\nl
7 & 3 36 16.7& -35 37 01.7 &21.01 & 1.23 &21.7 &1.31& 1376 $\pm84$ &1677 $\pm 150$\nl
8 & 3 36 15.4& -35 36 17.1 & \nodata & \nodata &22.3 &1.34& $z\simeq0.31$ &\nl
9 & 3 36 19.2& -35 36 28.7 &21.04 & 1.25 &21.8 &1.33& 1150 $\pm31$&1280 $\pm 150$\nl
10 & 3 36 21.5& -35 36 04.4&20.55 & 1.05 &21.4 &1.50& 815  $\pm30$&980 $\pm 150$\nl
11 & 3 36 25.0& -35 36 28.3&21.34 & 1.17 &22.2 &1.50& 1338 $\pm33$ &\nl
12 & 3 36 20.3& -35 35 15.3&21.97 & 0.94 &22.4 &1.06& 1736 $\pm31$&1701 $\pm 150$\nl
13 & 3 36 23.4& -35 35 37.2&21.51 & 0.91 &22.2 & \nodata  & 1247 $\pm30$&\nl
14 & 3 36 24.5& -35 35 36.8&21.17 & 1.37 &22.1 &1.65& 1260 $\pm66$ &\nl
15 & 3 36 23.2& -35 34 39.3&21.26 & 1.04 &22.0 &1.26& 1523 $\pm30$&\nl
16 & 3 36 25.0& -35 34 31.5&\nodata & \nodata &22.1 &1.56& $z\simeq0.27$& \nl
17 & 3 36 26.3& -35 34 20.7&21.55 & 1.14 &22.4 &1.33& 866  $\pm71$&\nl
18 & 3 36 31.4& -35 35 05.9&21.32 & 1.08 &22.3 &1.50& 1688 $\pm42$&\nl
19 & 3 36 34.5& -35 34 52.2&21.41 & 1.29 &22.3 & \nodata  & 1150 $\pm59$&\nl
20 & 3 36 28.5& -35 33 17.0&21.63 & 1.13 &22.3 &0.88& 1374 $\pm126$&\nl
21 & 3 36 35.7& -35 34 24.6&21.15 & 1.09 &22.0 &1.37& 1194 $\pm98$& 1062 $\pm 150$\nl
& & & & & & & & \nl
NGC 1399 (d) & 3 36 34.4 & -35 36 45.0 & 9.59 & 1.25 & \nodata & \nodata &
1447 $\pm 12$ & \nodata \nl
\enddata
\tablenotetext{(a)}{Magnitudes and colors from Kissler-Patig et al.~1997}
\tablenotetext{(b)}{Photographic measurements tabulated in Grillmair (1992)}
\tablenotetext{(c)}{Velocity from Grillmair et al.~(1994)}
\tablenotetext{(d)}{Data for NGC 1399 taken from de Vaucouleurs et al.~(1991),
except for V$-$I taken from Poulain (1988)}
\end{deluxetable}

\clearpage

\begin{deluxetable}{c r r r r r r r r r}
%\tablewidth{0pc}
\tablenum{2}
\tablecaption{Measured indices for the globular clusters}
\tablehead{
\colhead{ID} & \colhead{Mg2} & \colhead{MgH} & \colhead{Mgb} & \colhead{Fe5270}
& \colhead{Fe5335} & \colhead{H$\beta$} & \colhead{Gband} & \colhead{NaD} &
\colhead{TiO} \\
& $\pm 0.015$ & $\pm 0.015$ & $\pm 0.025$ & $\pm 0.020$ & $\pm 0.020$ &
$\pm 0.020$ & $\pm 0.045$ & $\pm 0.030$ & $\pm 0.030$ \\
}
\startdata
1 & 0.072 &0.005 &0.067 &0.052  &0.040 &0.079  &\nodata  &\nodata  &\nodata \\
2 & 0.319 &0.170 &0.214 &0.098  &0.091 &0.115  &\nodata  &0.154    &0.030 \\
3 & 0.111 &0.003 &0.089 &0.035  &0.022 &0.058  &0.060    &\nodata  &\nodata \\
5 & 0.094 &0.020 &0.071 &0.066  &0.059 &0.082  &0.096    &0.042    &0.033 \\
6 & 0.129 &0.038 &0.127 &0.021  &0.041 &0.088  &\nodata  &\nodata  &0.026 \\
7 & 0.228 &0.078 &0.182 &0.053  &0.090 &0.060  &0.114    &0.174    &0.028 \\
9 & 0.174 &0.059 &0.123 &0.068  &0.072 &0.080  &0.136    &0.063    &0.027 \\
10 & 0.066 &0.022 &0.034 &0.028  &0.037 &0.098  &0.081    &\nodata  &0.017 \\
11 & 0.179 &0.051 &0.129 &0.061  &0.056 &0.100  &\nodata  &\nodata  &0.029 \\
12 & 0.032 &0.040 &0.076 &0.045  &0.023 &0.114  &0.079    &0.065    &\nodata \\
13 & 0.066 &0.016 &0.067 &0.044  &0.008 &0.108  &0.096    &\nodata  &0.017 \\
14 & 0.339 &0.153 &0.227 &0.099  &0.083 &0.080  &\nodata  &0.283    &0.034 \\
15 & 0.114 &0.026 &0.097 &0.064  &0.051 &0.088  &0.103    &\nodata  &0.023 \\
17 & 0.210 &0.058 &0.160 &0.044  &0.048 &0.102  &0.169    &0.071    &0.036 \\
18 & 0.181 &0.050 &0.156 &0.079  &0.067 &0.087  &\nodata  &\nodata  &0.039 \\
19 & 0.244 &0.078 &0.156 &0.068  &0.085 &0.084  &\nodata  &0.168    &0.062 \\
20 & 0.122 &0.032 &0.101 &0.048  &0.055 &0.082  &\nodata  &\nodata  &0.048 \\
21 & 0.168 &0.051 &0.129 &0.077  &0.040 &0.078  &0.182    &\nodata  &0.055 \\
& & & & & & & & & \\
NGC 1399 (a) & 0.371 & 0.201 & 0.202 & 0.081 & 0.082 & 0.052 & 0.197 & 0.261 & 
0.059 \\ 
& $\pm0.008$ & $\pm0.007$ & $\pm0.012$ & $\pm0.007$ & $\pm0.014$
& $\pm0.013$ & $\pm0.039$ & $\pm0.013$ & $\pm0.007$ \\
NGC 1399 (b) & 0.288 & 0.186 & 0.113 & 0.038 & \nodata & 0.045 & 0.223 
& 0.122 & \nodata \\ 
& $\pm0.024$ & $\pm0.021$ & $\pm0.040$ & $\pm0.029$ & \nodata
& $\pm0.037$ & $\pm0.052$ & $\pm0.038$ & \nodata \\
\enddata
\tablenotetext{}{All bandpass definition were taken from Brodie \& Huchra (1990)
}
\tablenotetext{(a)}{Taken from Trager (1997)}
\tablenotetext{(b)}{Taken from Huchra et al.~(1996), {\bf note that,
contrary to Trager's data, the measurements were made on the galaxy spectra
without correcting for velocity dispersion}, and can therefore not be
directly compared with them or the globular cluster measurements}
\end{deluxetable}

\clearpage

\begin{deluxetable}{c rrrrrrrr cc}
%\tablewidth{0pc}
\tablenum{3}
\tablecaption{$[$Fe/H$]$ calculated from various line indices for the globular
clusters}
\tablehead{
\colhead{ID} & \colhead{  Mg2} & \colhead{  Fe5270} & \colhead{ 
MgH} & \colhead{  Mgb} & \colhead{  Gband} & \colhead{NaD} &
\colhead{  CN} & \colhead{  H+K} & \colhead{mean from} & \colhead{weighted} \\
&  &  &  &  &  &  &  &  & \colhead{Mg2 Fe5270(a)} & \colhead{mean(a)}\\
& $\pm 0.15$ & $\pm 0.35$ & $\pm 0.30$ & $\pm 0.35$ & $\pm 0.50$ & $\pm
0.40$ & $\pm 0.50$ & $\pm 0.50$ & \colhead{$\pm 0.20$} & \colhead{$\pm 0.20$} \\
}
\startdata
1 &-1.51& -1.04& -1.74& -1.27& \nodata& \nodata& -1.19& \nodata& -1.27& -1.30\\
2 &  0.18&   -0.10&  0.41&  0.80& \nodata& -0.14& \nodata& \nodata&  0.04&  0.12\\
3 & -1.11&  -1.37& -1.78&  -0.96& -1.76& \nodata& -1.17& \nodata& -1.24& -1.29\\
5 & -1.28&  -0.75& -1.42& -1.21& -1.36&-1.75& -0.95& -1.13& -1.01& -1.12\\
6 &-0.94&  -1.65& -1.05& -0.42& \nodata& \nodata& \nodata& \nodata& -1.29& -1.20\\
7 & -0.27&  -1.01&  -0.23&  0.35& -1.16&  0.15& -1.03& \nodata& -0.64& -0.55\\
9 &-0.48&  -0.71&  -0.63&  -0.49&  -0.90& \nodata& -0.45& \nodata& -0.59& -0.60\\
10 &-1.56&  -1.52&   -1.40&  -1.74& -1.53& \nodata& -1.35& \nodata& -1.54& -1.53\\
11 & -0.53&  -0.85& -0.79&   -0.4& \nodata& \nodata& \nodata& \nodata& -0.69& -0.67\\
12 &-1.90&  -1.18&  -1.03& -1.13& -1.55&-1.43& \nodata& \nodata& -1.54& -1.47\\
13 &-1.56&  -1.19& -1.51& -1.26& -1.35& \nodata& -1.42& -1.24& -1.37& -1.37\\
14 &  0.26& -0.08&  0.26&  0.99& \nodata& 1.72& \nodata& \nodata&  0.09&  0.30\\
15 &-1.09&  -0.78&   -1.30&  -0.85& -1.28& \nodata& \nodata& \nodata& -0.93& -0.98\\
17 &-0.38& -1.21&  -0.65&   0.04& -0.52& -1.33& -0.73& -0.89& -0.79& -0.81\\
18 &-0.42& -0.50&  -0.81&  -0.01& \nodata& \nodata& \nodata& \nodata& -0.46& -0.45\\
19 & -0.17&  -0.72& -0.25&  -0.02& \nodata&  0.07& \nodata& \nodata& -0.40& -0.32\\
20 & -1.00&  -1.12& -1.19&  -0.79& \nodata& \nodata& -0.99& \nodata& -1.06& -1.04\\
21 & -0.54&  -0.53&  -0.79& -0.38& -0.39& \nodata& \nodata& \nodata& -0.54& -0.54\\
& & & & & & & & & & \\
NGC 1399 (b) & 0.52 & -0.44 & 0.85 & 1.19 & -0.21 & 1.41 & \nodata& \nodata&
0.04 & 0.26 \\
\enddata
\tablenotetext{}{All [Fe/H] values were calculated using the 
index--metallicity relations of Brodie \& Huchra (1990)}
\tablenotetext{(a)}{
The last two columns list the arithmetic mean of the metallicity derived 
from Mg2 and Fe5370 and a weighted mean (see text for assigned weights) 
respectively}
\tablenotetext{(b)}{calculated from the values in Table 2}
\end{deluxetable}

\clearpage

%%%%%%%%%%%%%%%%%%%%%  the velocity / position plots  %%%%%%%%%%%%%%%%%%%

\begin{figure}
\psfig{figure=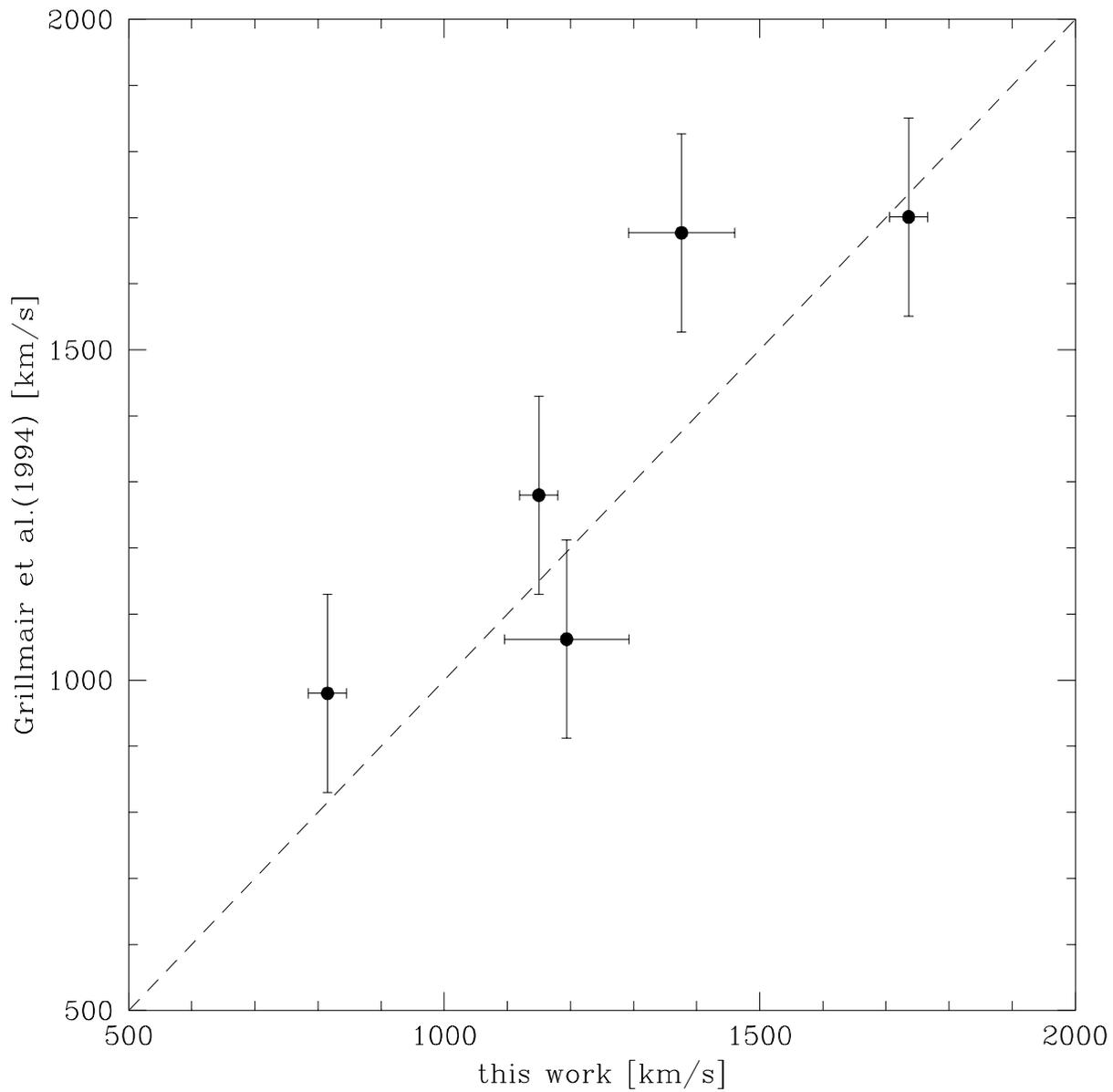,height=16cm,width=16cm
,bbllx=8mm,bblly=57mm,bburx=205mm,bbury=245mm}
\caption{
%\figcaption[Kissler-Patig.fig1.ps]{
Comparison between velocities derived in this work and in Grillmair et 
al.~(1994) for the five globular clusters in common to both samples
}
\end {figure}

\clearpage

\begin{figure}
\psfig{figure=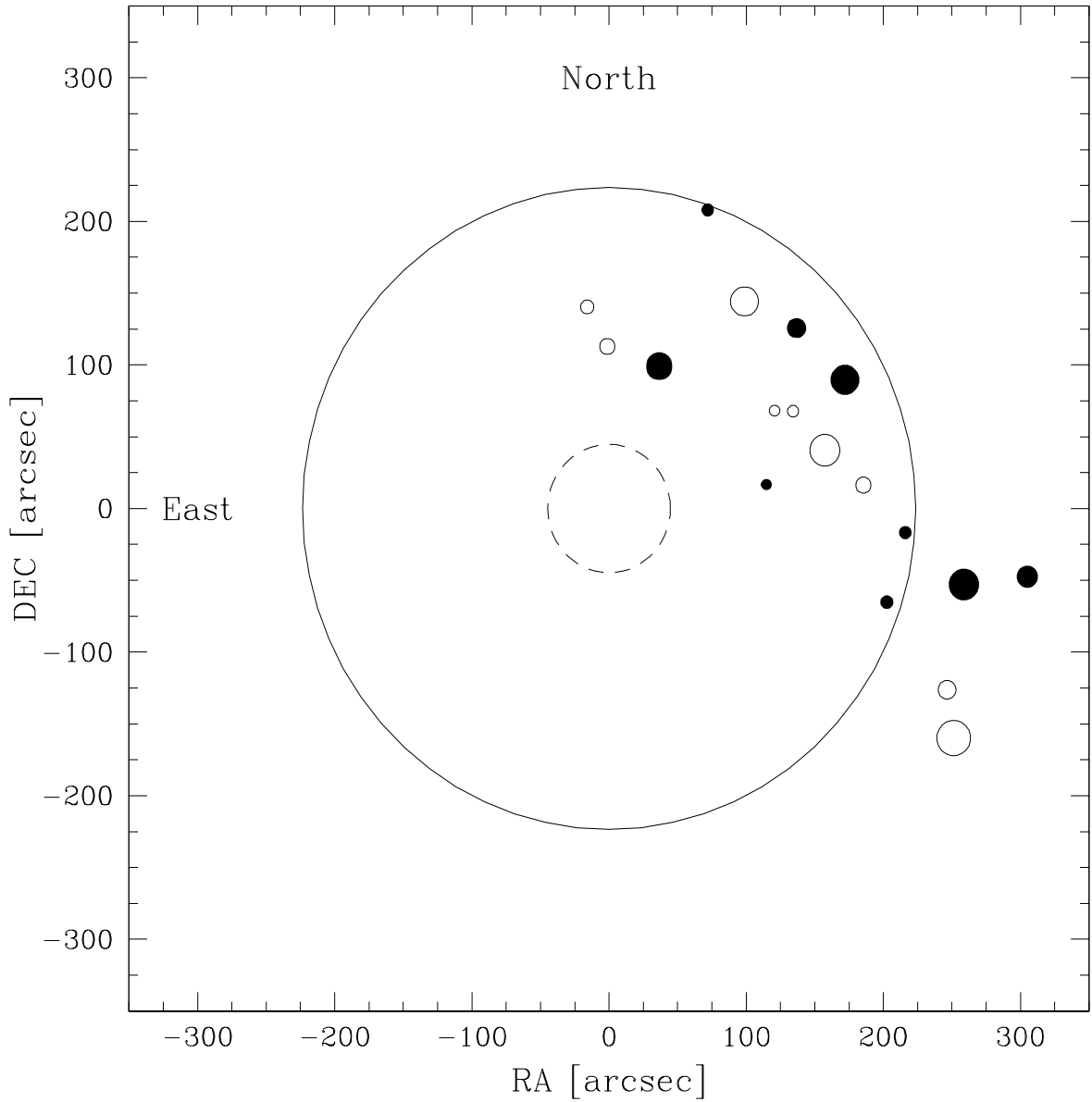,height=16cm,width=16cm
,bbllx=8mm,bblly=57mm,bburx=205mm,bbury=245mm}
\caption{
%\figcaption[Kissler-Patig.fig2.ps]{
Position of approaching (open) and receding (solid) globular clusters 
(reference velocity is 1293 km/s), the rings mark 1 and 5 r$_{\rm eff}$
around NGC 1399 
}
\end {figure}

\clearpage

\begin{figure}
\psfig{figure=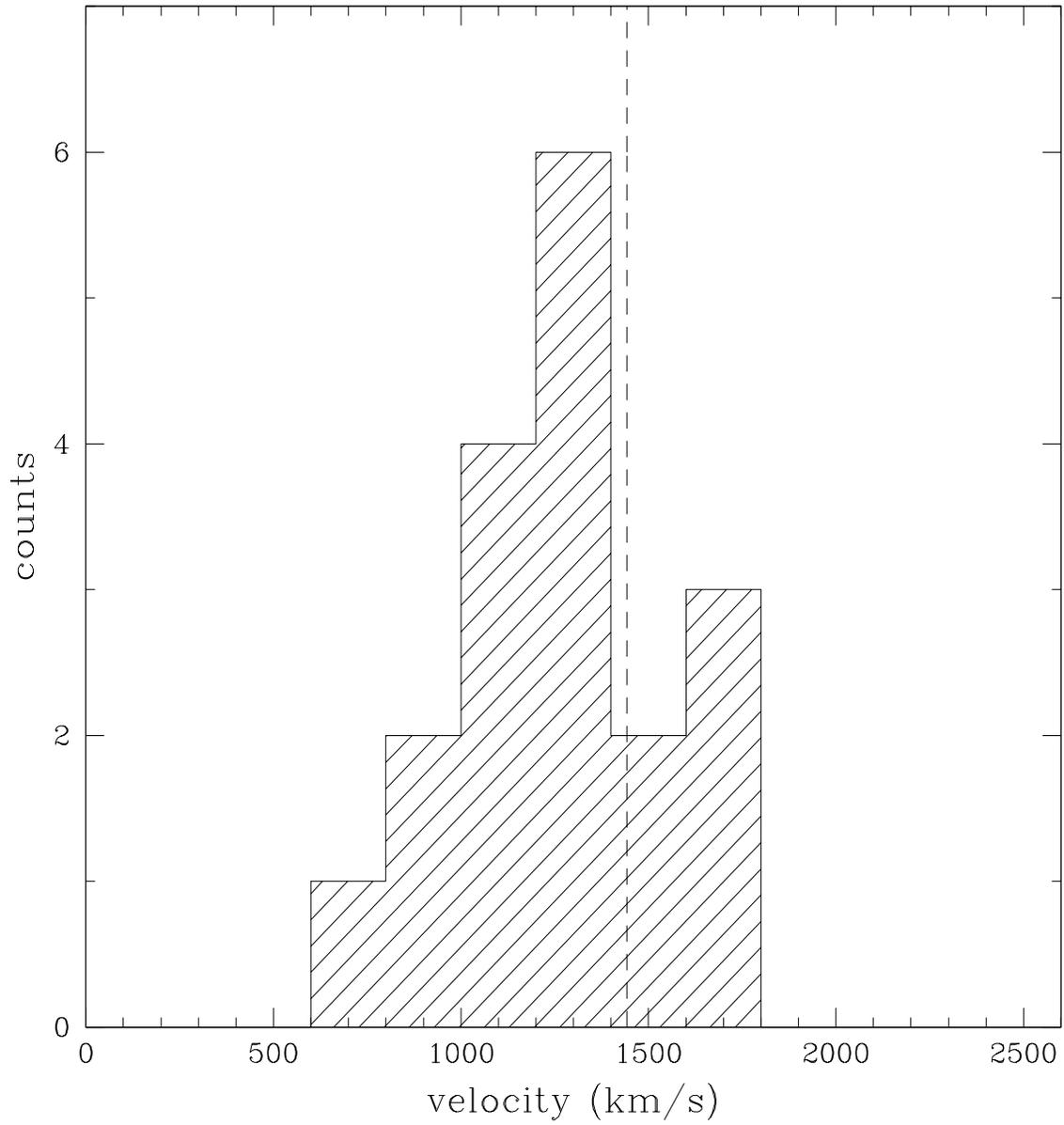,height=16cm,width=16cm
,bbllx=8mm,bblly=57mm,bburx=205mm,bbury=245mm}
\caption{
%\figcaption[Kissler-Patig.fig3.ps]{
Velocity distribution of our 18 globular clusters around NGC
1399. The dashed line marks the radial velocity of the stellar component
(de Vaucouleurs et al.~1991)
}
\end {figure}

\clearpage

\begin{figure}
\psfig{figure=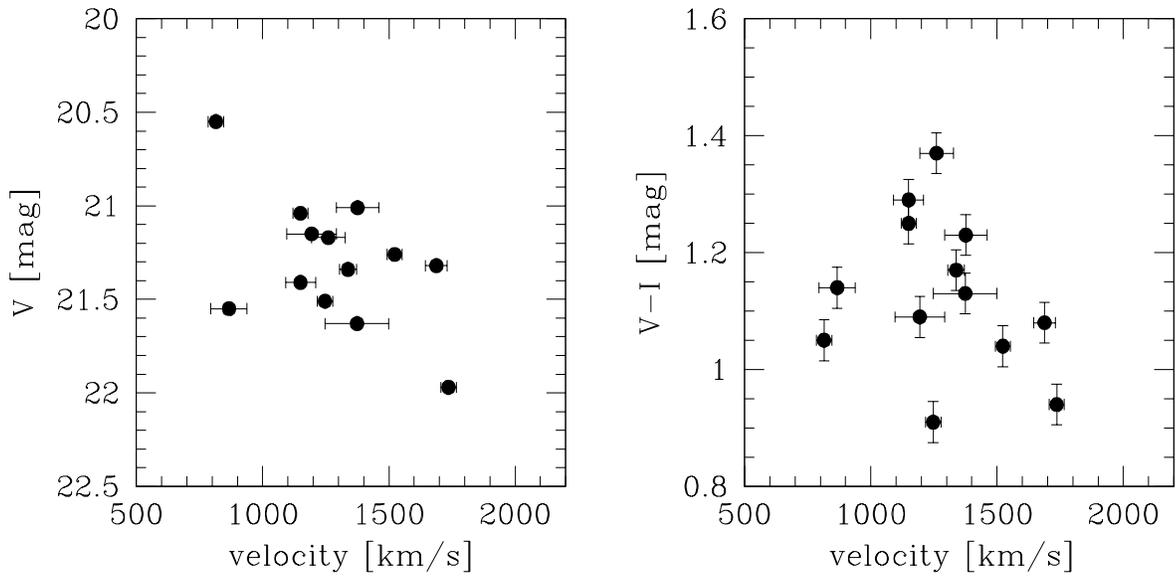,height=16cm,width=16cm
,bbllx=8mm,bblly=57mm,bburx=205mm,bbury=245mm}
\caption{
%\figcaption[Kissler-Patig.fig4.ps]{
Velocities versus V magnitudes and V$-$I colors of the globular clusters.
No trend can be seen in our sample
}
\end {figure}

\clearpage

\begin{figure}
\psfig{figure=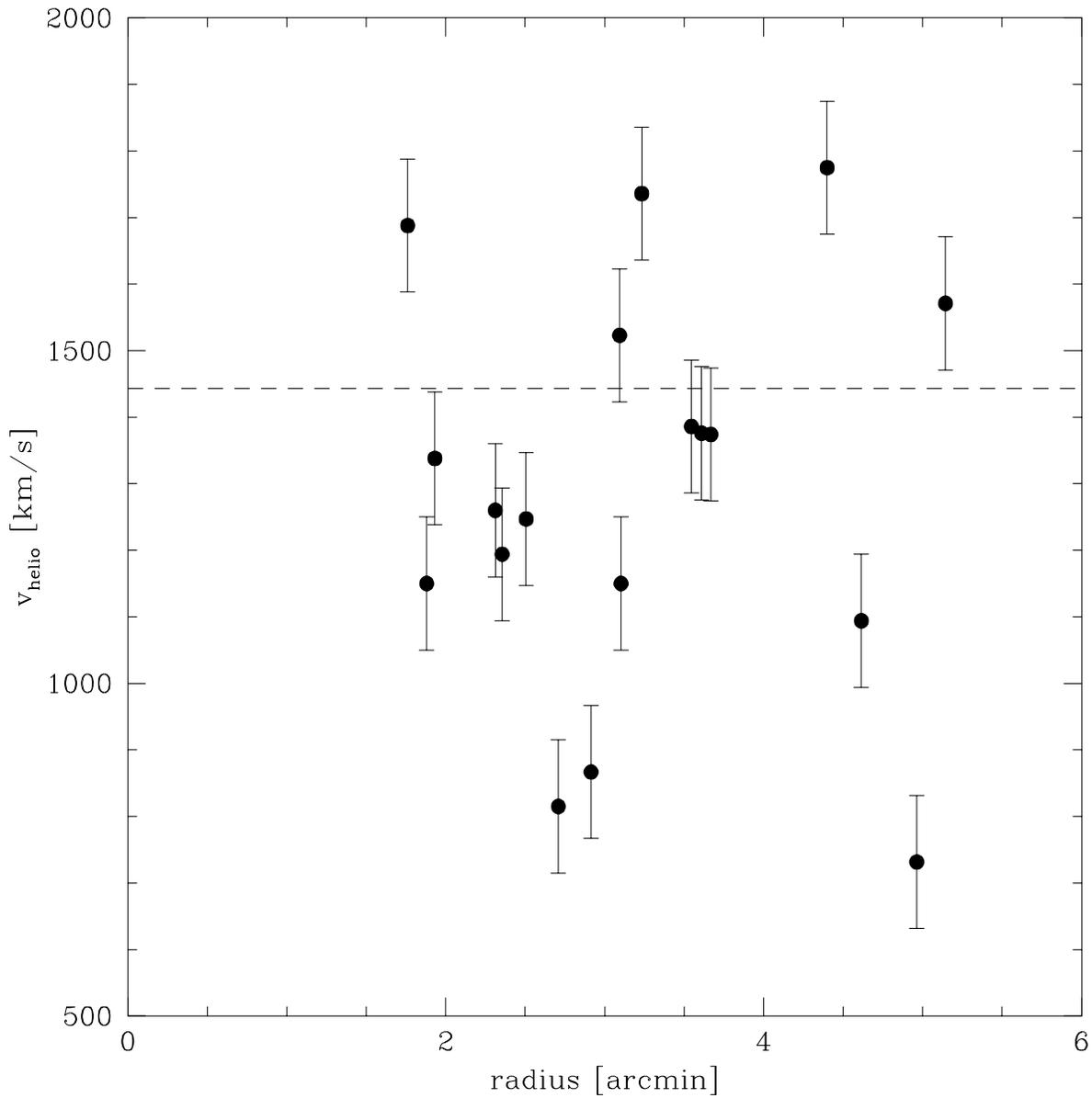,height=16cm,width=16cm
,bbllx=8mm,bblly=57mm,bburx=205mm,bbury=245mm}
\caption{
%\figcaption[Kissler-Patig.fig5.ps]{
Velocities versus radius (in arcmin) for the globular clusters. The dashed
line marks the radial velocity of the stellar component (taken from de
Vaucouleurs et al.~1991). No trend can be seen in our sample
}
\end {figure}

%%%%%%%%%%%%%%%%%%%%%%%%%%%%%%% the Abundances  plots  %%%%%%%%%%%%%%%%%%%%%%%%%

\clearpage

\begin{figure}
\psfig{figure=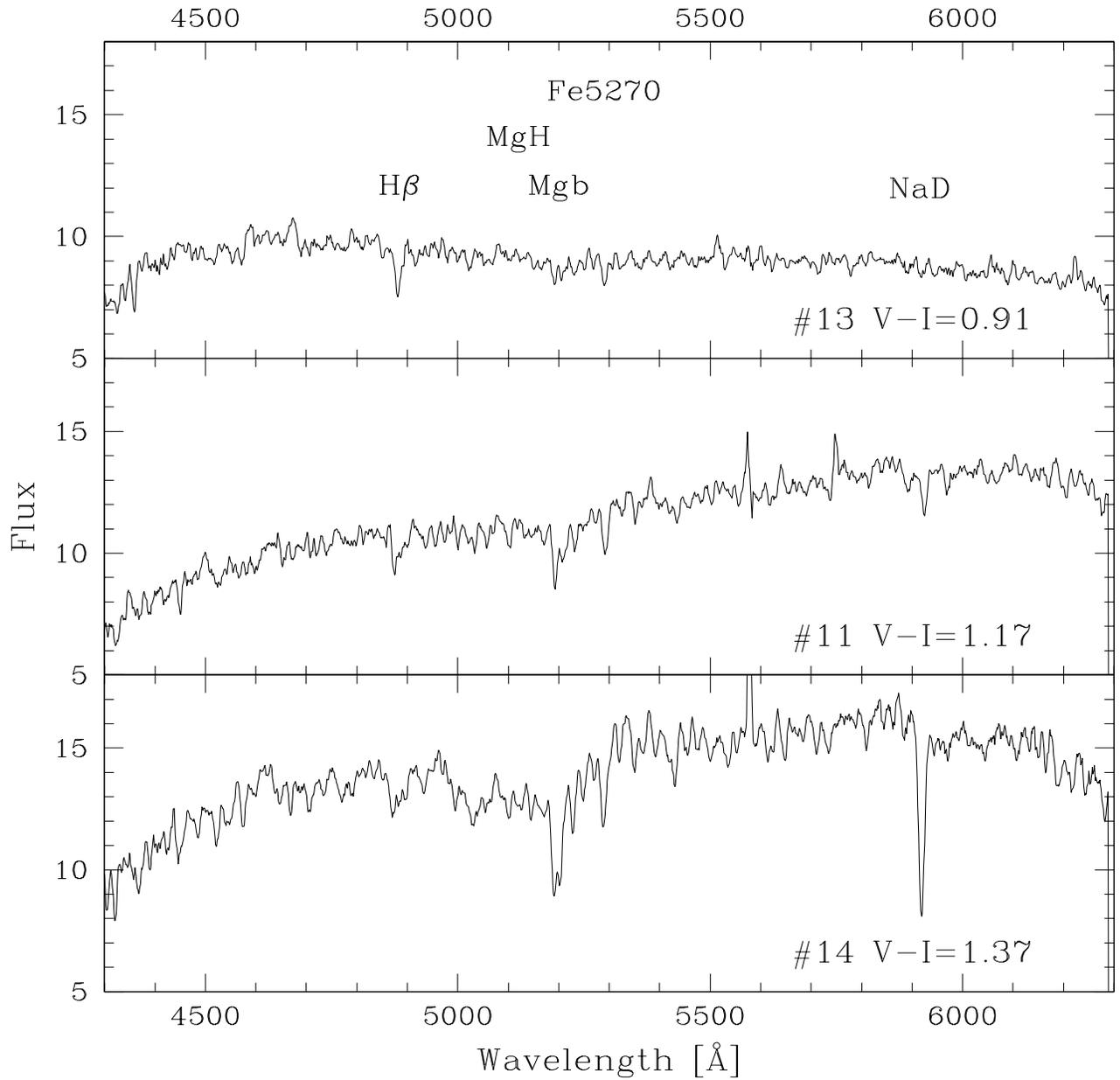,height=16cm,width=16cm
,bbllx=8mm,bblly=57mm,bburx=185mm,bbury=245mm}
\caption{
%\figcaption[Kissler-Patig.fig6.ps]{
Three representative spectra, ranging from blue, over
red, to very red color. The spectra were smoothed over 3 pixels. While 
the H$\beta$ gets slightly weaker from the blue to the red object, the metal 
lines (Mg, Fe, Na) become much stronger
}
\end {figure}

\clearpage

\begin{figure}
\psfig{figure=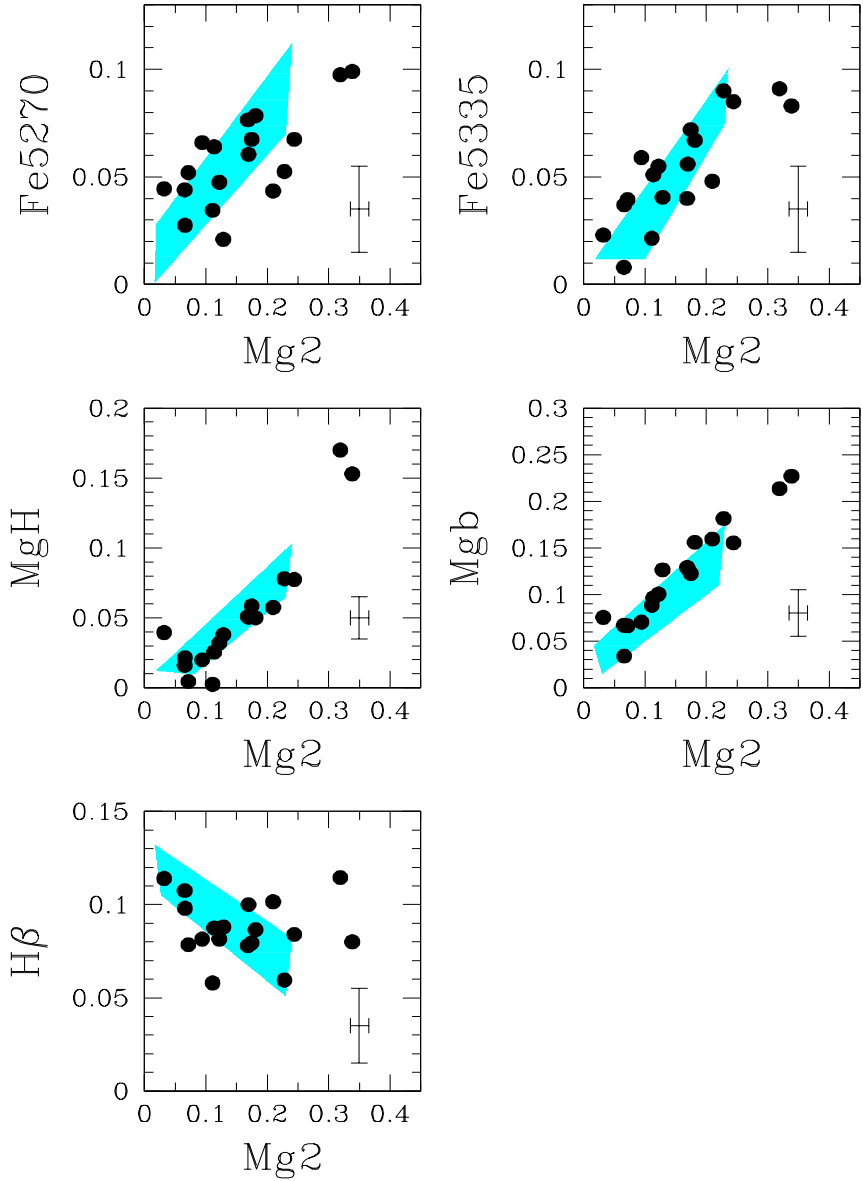,height=16cm,width=16cm
,bbllx=8mm,bblly=57mm,bburx=185mm,bbury=245mm}
\caption{
%\figcaption[Kissler-Patig.fig7.ps]{
Mg2 versus various other indices, with the range spanned by Milky Way and
M31 globular clusters shown as shaded region (taken from Brodie \& Huchra
1990 and Burstein et al.~1984). Beside two very metal--rich
objects, all globular clusters are consistent with abundances found in the
Milky Way and M31 globular clusters
}
\end {figure}

\clearpage

\begin{figure}
\psfig{figure=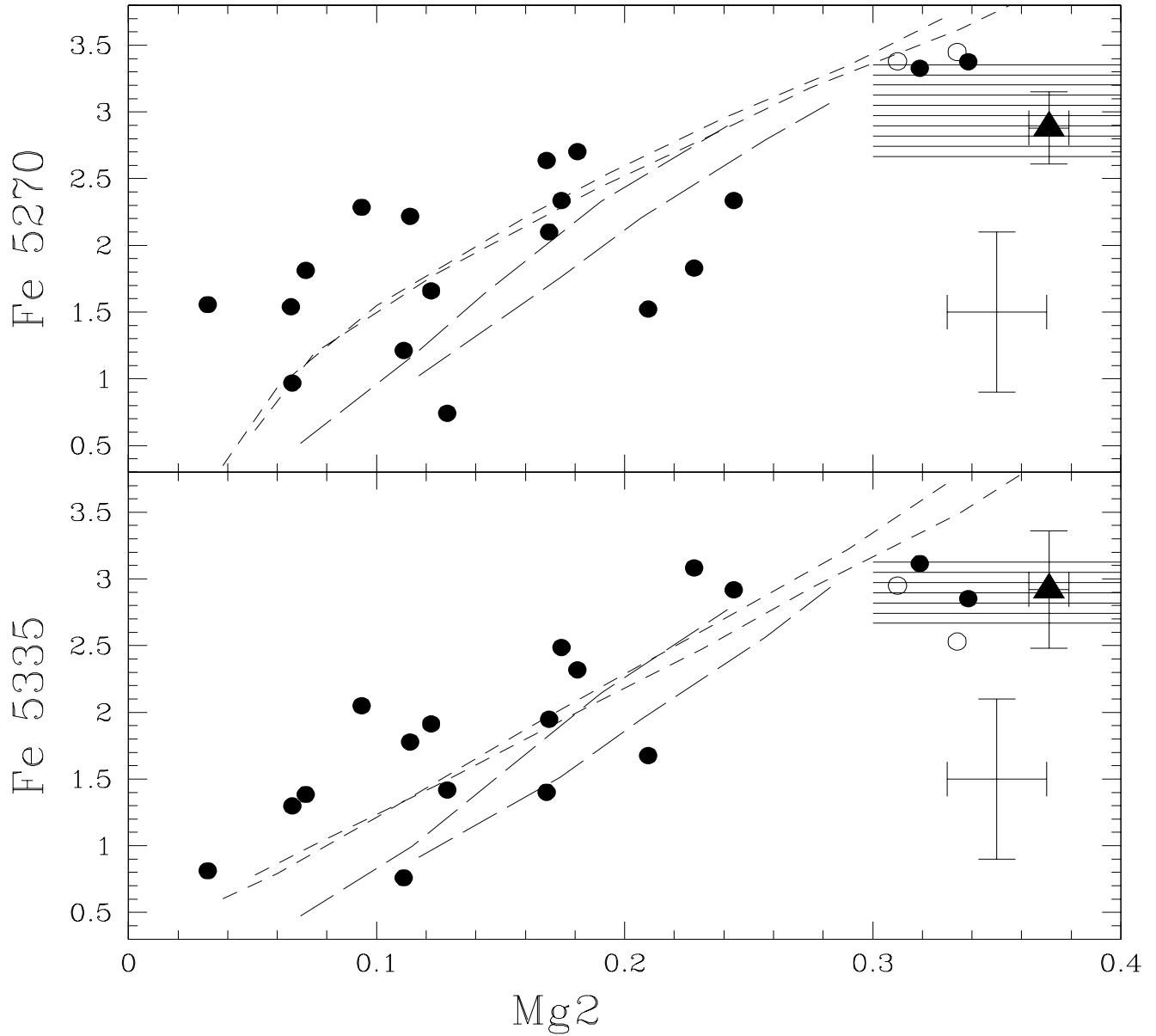,height=16cm,width=16cm
,bbllx=8mm,bblly=57mm,bburx=185mm,bbury=245mm}
\caption{
%\figcaption[Kissler-Patig.fig8.ps]{
Mg2 versus the equivalent width of Fe5270 and Fe5335. Tracks for 8 and 16
Gyr old stellar populations (metallicity varying between Z=0.001 and
Z=0.04) from Fritze-v.~Alvensleben \& Burkert (1995) are plotted as long dashed lines,
tracks for 8 and 17 Gyr old stellar populations (metallicity varying between
$[$Fe/H$]=-2.0$ and $0.5$ dex) from Worthey (1994) are plotted as short
dashed lines. The regions spanned by elliptical galaxies in Worthey et
al.~(1992) are lined. NGC 1399
(taken from Trager 1997) is plotted as a triangle. Open circles are
measurements using the updated Lick/IDS bandpasses on artificially degraded
spectra (see text)
}
\end {figure}

\clearpage
 
\begin{figure}
\psfig{figure=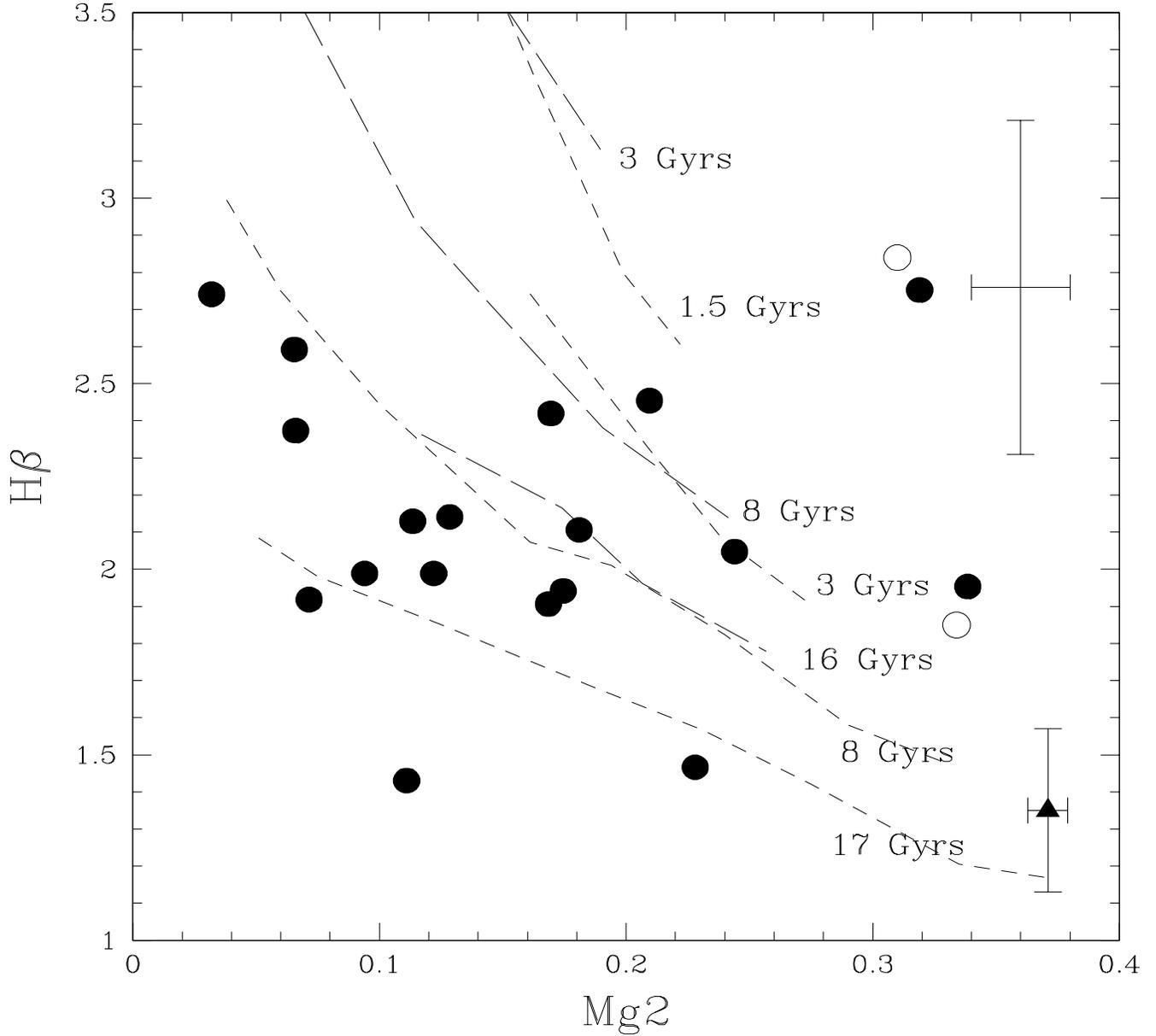,height=16cm,width=16cm
,bbllx=8mm,bblly=57mm,bburx=185mm,bbury=245mm}
\caption{
%\figcaption[Kissler-Patig.fig9.ps]{
H$\beta$ (here in \AA ) versus Mg2 with tracks from Fritze-v.~Alvensleben 
\& Burkert (1995) 
(long dashed lines, metallicity varying between Z=0.001 and Z=0.04) and from 
Worthey (1994) (short dashed lines, metallicity varying between
$[$Fe/H$]=-2.0$ and $0.5$ dex). Other symbols as in Fig.~8
}
\end {figure}

\clearpage

%%%%%%%%%%%%%%%%%%%%%%%%%%%%%% the Fe/H plots %%%%%%%%%%%%%%%%%%%%%%%%%%%%%%%%

\begin{figure}
\psfig{figure=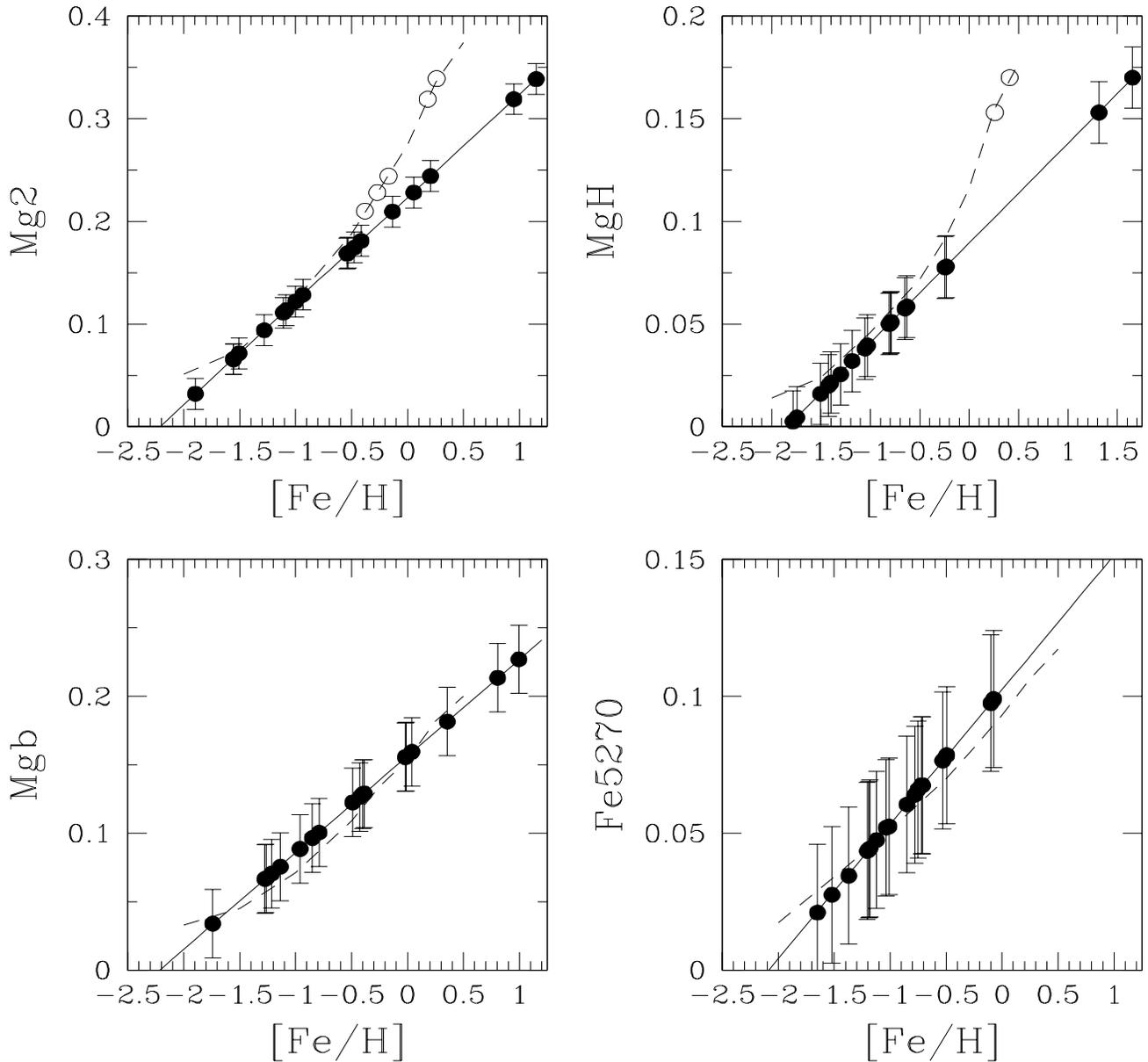,height=16cm,width=16cm
,bbllx=8mm,bblly=57mm,bburx=185mm,bbury=245mm}
\caption{
%\figcaption[Kissler-Patig.fig10.ps]{
Comparison between the metallicities derived from Mg2, MgH, Mgb and Fe5270
using the relation of Brodie \& Huchra (1990) (solid line), and the values 
predicted by the models of Worthey (1994) (dashed lines). The solid dots show 
our data points, when using the Brodie \& Huchra (1990) relation. The open
dots show our data points once corrected for the non--linear behaviour of
Mg at high metallicities. We corrected five values derived from Mg2,
two derived from MgH. Values derived from Mgb or Fe5270 do not need any
corrections.
}
\end {figure}

\clearpage

\begin{figure}
\psfig{figure=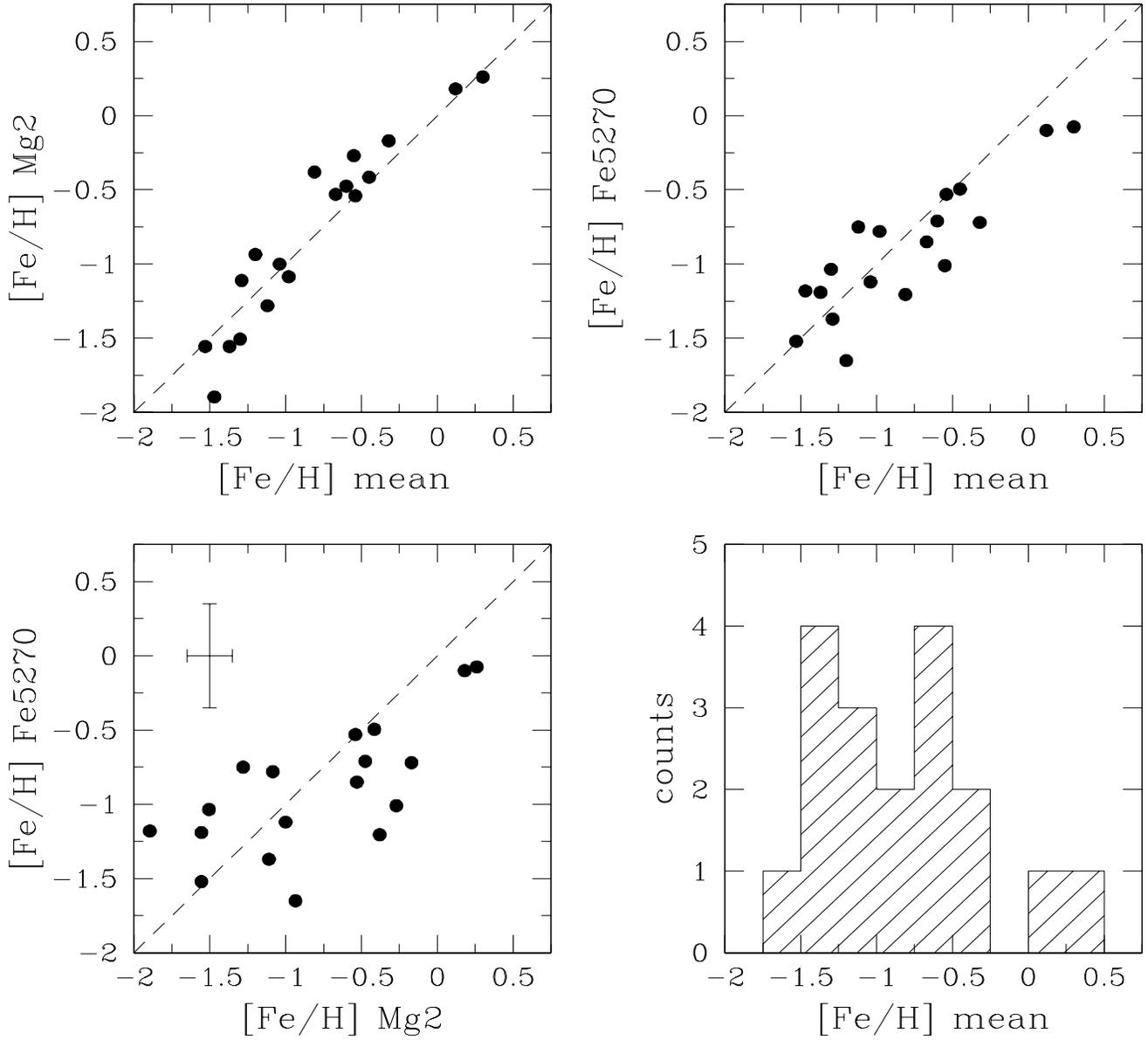,height=16cm,width=16cm
,bbllx=8mm,bblly=57mm,bburx=185mm,bbury=245mm}
\caption{
%\figcaption[Kissler-Patig.fig11.ps]{
Comparison between the metallicities derived from Mg2, Fe5270, and mean
metallicity as defined in the text. The lower right panel shows a histogram
over the mean metallicities of the globular clusters
}
\end {figure}

%%%%%%%%%%%%%%%%%%%%%%%%%%%%%% the V-I  plots  %%%%%%%%%%%%%%%%%%%%%%%%%

\clearpage

\begin{figure}
\psfig{figure=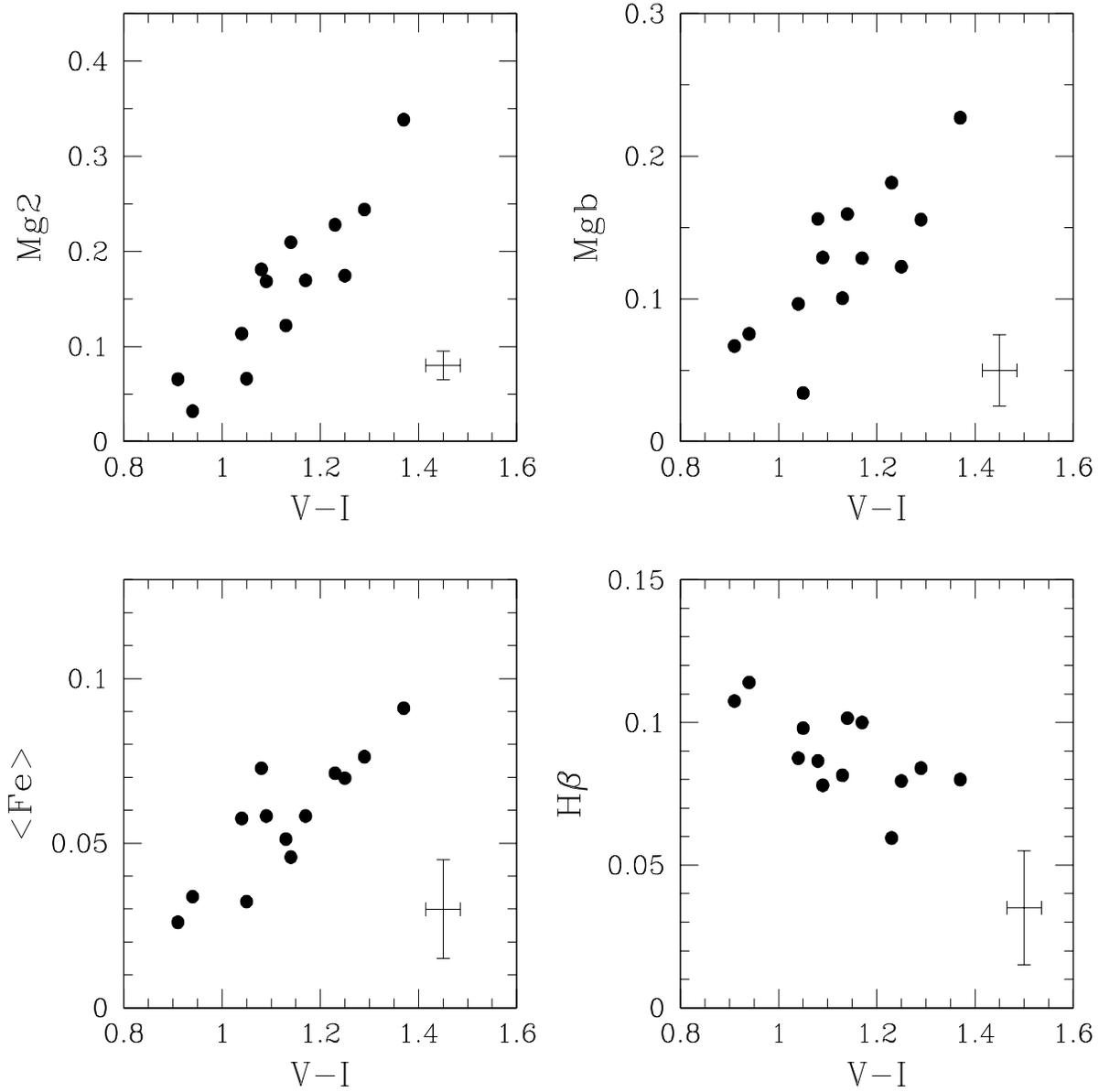,height=16cm,width=16cm
,bbllx=8mm,bblly=57mm,bburx=205mm,bbury=245mm}
\caption{
%\figcaption[Kissler-Patig.fig12.ps]{
V$-$I colors of the globular clusters versus Mg, Fe and H$\beta$. The
broad--band color correlates well both with the metal and (inverse)
Hydrogen abundance 
}
\end {figure}

\clearpage

\begin{figure}
\psfig{figure=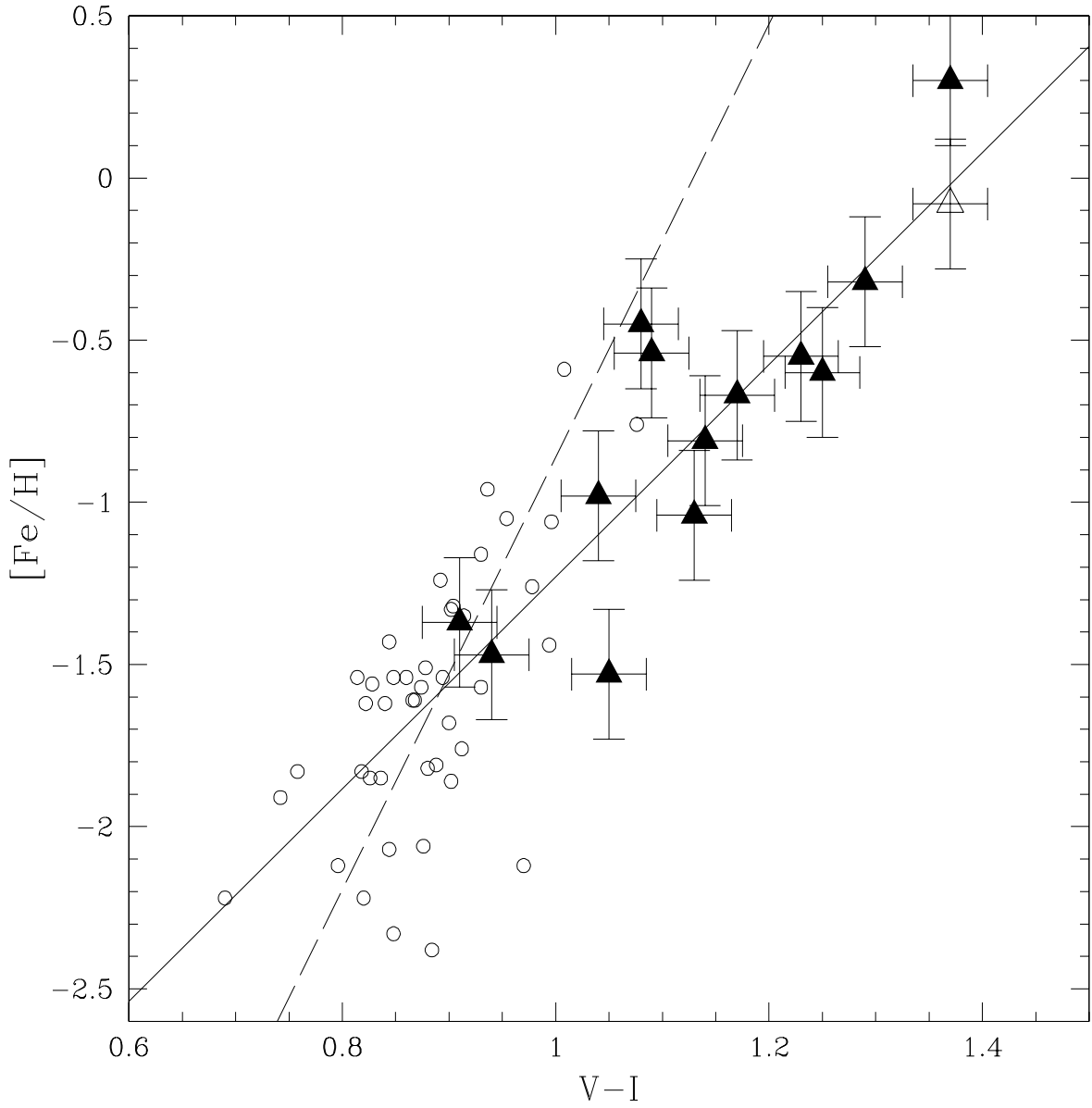,height=16cm,width=16cm
,bbllx=8mm,bblly=57mm,bburx=205mm,bbury=245mm}
\caption{
%\figcaption[Kissler-Patig.fig13.ps]{
V$-$I color versus mean metallicity of the globular clusters in NGC 1399
(triangles), for cluster \#14 the metallicity from Fe5270 alone is also shown 
(open triangle). Over--plotted as open circles are all Milky Way globular
clusters with E(B$-$V)$<0.2$ (taken from Harris 1996, de--reddened according
to Rieke \& Lebofsky 1985). The solid line is the best fit to the data, the
dashed line was derived from the Milky Way sample by Kissler-Patig et
al.~(1997) and extrapolated to higher metallicities
}
\end {figure}

\end{document}